\documentclass[%
 aip,
jcp,
tightenlines,
runinaddress,
 amsmath,amssymb,
]{revtex4-1}
\usepackage{graphicx}
\usepackage{dcolumn}
\usepackage{bm}

\usepackage{subcaption}
\usepackage[para]{threeparttable}
\usepackage{natbib}
\begin{document}


\title[]{Electrostatic solvation free energies of charged hard spheres using molecular dynamics with density functional theory interactions}

\author{Timothy T. Duignan}
 \email{timothy.duignan@pnnl.gov}
 \author{Marcel D. Baer}
\author{Gregory K. Schenter}
\author{Chistopher J. Mundy}
 \email{chris.mundy@pnnl.gov}%
 \affiliation{Physical Science Division, Pacific Northwest National Laboratory, P.O. Box 999,Richland, Washington 99352, USA}
\date{\today}

\begin{abstract}
Determining the solvation free energies of single ions in water is one of the most fundamental problems in physical chemistry and yet many unresolved questions remain.  In particular, the ability to decompose the solvation free energy into simple and intuitive contributions will have important implications for models of electrolyte solution.  Here, we provide  definitions of the various types of single ion solvation free energies based on different simulation protocols. We calculate solvation free energies of charged hard spheres using density functional theory interaction potentials with molecular dynamics simulation  (DFT-MD) and isolate the effects of charge and cavitation, comparing to the Born (linear response) model.  We show that using uncorrected Ewald summation leads to unphysical values for the single ion solvation free energy and that charging free energies for cations are approximately linear as a function of charge but that there is a small non-linearity for small anions. The charge hydration asymmetry (CHA) for hard spheres, determined with quantum mechanics, is much larger than for the analogous real ions. This suggests that real ions, particularly anions, are significantly more complex than simple charged hard spheres, a commonly employed representation. 
\end{abstract}

\maketitle

\section{Introduction}
There is no consensus in the literature regarding the solvation free energies of individual ions in water.\cite{Hunenberger2011,Vlcek2015,Pollard2016}  Estimates of these quantities are spread over a range of 50 kJmol$^{-1}$. In addition to this, the size of the various contributions to these solvation free energies remain largely unknown. 
This significantly hinders our understanding of the many important systems where electrolyte solutions play a central role because knowledge of these values is vital for testing and parametrizing theoretical models of electrolyte solutions. \cite{Hunenberger2011,Horinek2009,Grossfield2003}
Surprisingly, only a single  study of the solvation free energy of ions in water using density functional theory interaction potentials with molecular dynamics simulation (DFT-MD) has been published.\cite{Leung2009}  
Rather than tackling the solvation free energy of real ions in water directly, it is  useful to consider the solvation free energy of the simplest possible model of an ion: a charged hard sphere.  The solvation free energy of a charged hard sphere can be exactly partitioned into a cavity formation free energy and an electrostatic charging free energy. It is useful to study this simple model of an ion as the solvation free energies of real ions include a significant contribution from quantum mechanical interactions such as dispersion and exchange terms. The role of the pure electrostatic interaction is therefore obscured by these terms, which prevents physical insight.  In addition these interactions are necessarily treated at a highly approximate level with classical-MD and continuum solvent models. By calculating the solvation free energy of a charged hard sphere these terms are excluded and a more direct comparison with these simpler models is possible allowing us to test and parameterize them. Any differences between classical-MD and DFT-MD in the solvation free energies of charged hard spheres cannot be attributed to issues with the Lennard Jones parameters but must be ascribable to the incorrect response of the water to the electric field of the ion. 
An example of this is the fundamental concept of the charge hydration asymmetry (CHA).\cite{Mukhopadhyay2012} The CHA refers to the different response of  water to a positive charge versus a negative charge of the same size and it is important that  models of water accurately reproduce it. Unfortunately,  we still have very little idea of how large this asymmetry is for real water. This is confounded in real systems because of the quantum mechanical interactions with the ion, namely dispersion and exchange interactions, which obscure the size of the CHA.  This quantity can only be properly determined using the framework of quantum mechanical simulation of charged hard spheres in water, \emph{i.e.}, DFT-MD. Studying these model ions also affords an unambiguous route to examine the linear response behavior of \emph{ab initio} water allowing direct comparison to the Born model and other reduced treatments of electrolyte properties.  

Before we can answer the aforementioned questions there is a more fundamental issue which must be resolved that provides an additional motivation for studying charged hard spheres rather than real ions. There is substantial confusion and debate in the literature regarding the very definition of single ion solvation free energies. This stems from the fact that single ion solvation free energies rely on a reference for the zero of the electrostatic potential. If two different but correct methods assume a different zero, then the single ion solvation free energies will differ by $q\Delta\phi$. This ambiguity is not present for  cation-anion pairs due to the obvious cancellation. Moreover, the precise zero of the electrostatic potential for a given theoretical or experimental method is often unclear as it can depend sensitively on the simulation protocol or other assumptions. This problem is amplified due to a lack of uniformity in the literature concerning terminology and notation for the relevant quantities.  This confusion is the main reason for the large spread in estimates of these quantities in the literature. 

Many ``extra thermodynamic assumptions'' have been used to try and determine single ion solvation free energies.  The cluster pair approximation (CPA) and the tetraphenylarsonium tetraphenylborate (TATB) assumption  are the two most well known examples. The first uses the formation free energies of gas phase ion-water  clusters to determine a value\cite{tissandier1998,Kelly2006} and the second assumes the equivalence of two large hydrophobic ions to evenly split the solvation free energy.\cite{Marcus2000}  
Unfortunately, these efforts have not resolved the issue.\cite{Vlcek2013,Pollard2014,Schurhammer2000,Scheu2014} 

Purely theoretical methods do not give consistent values for these quantities. For instance, one common approach is to calculate the free energy of forming small ion-water clusters and solvating them in a dielectric continuum.\cite{Zhan2001,Asthagiri2005,Bryantsev2008,Sabo2013} One issue is that it not obvious what the zero of the electrostatic potential is with these calculations.
 Another is that the solvation free energies determined with these methods differ by 50 kJ mol$^{-1}$ or more depending on what thermodynamic cycle is used and there is debate about what the best cycle is.\cite{Bryantsev2008,Merchant2011,Rogers2011,Sabo2013}  If a water cluster cycle is used, agreement with the CPA based values is achieved for both cations and anions.\cite{Bryantsev2008,Zhan2001,Zhan2004} 
An alternative cycle treats the water molecules individually rather than as a cluster when determining their desolvation energy. This approach relies on knowing a coordination number, and it results in values that  agree with Marcus' solvation free energies for cations\cite{Asthagiri2003,Asthagiri2004,Chaudhari2015d} but for the hydroxide anion the solvation free energy differs by approximately 50 kJ mol$^{-1}$ from Marcus' value.\cite{Asthagiri2003} Ref.~\citenum{Asthagiri2003} also reports single ion solvation free energies calculated by inserting a cluster into explicit solvent rather than into a dielectric continuum. However, because these calculations use Ewald summation they need to be corrected as discussed below.

Classical-MD appears to show significant model dependence in both surface potentials\cite{Remsing2014} and solvation free energies.\cite{Horinek2009} This is not necessarily surprising;  these models are mainly parametrized and compared against bulk properties of aqueous solutions and so they are not necessarily reliable in the highly asymmetric environment of the air-water interface. In addition, the simple functional forms used for the interaction potentials must describe both classical electrostatic and quantum mechanical interactions, which is difficult to achieve. 

In this work we aim to use state-of-the-art DFT-MD techniques to establish the electrostatic solvation free energies of charged hard spheres in water and compare with the Born model and with classical-MD. We investigate monovalent charged hard spheres of the size relevant to small monatomic ions and find the CHA for DFT water is significantly larger than that obtained for water modeled with classical-MD. This finding points to an oversimplification of classical-MD. A mapping between different definitions of solvation free energies is constructed allowing comparison to other definitions and notations in the literature. 
 
\section{Theory and Definitions}
\subsection{Solvation Free Energies}
The key quantity that we need to calculate is the  `excess chemical potential' of an ion X in solution given by:
\begin{equation}
\mu^*_{X}=-k_{\text{B}}T\ln\left< e ^{-\beta U_{XS}}\right>_{0}-E^\text{Vac}_X
\label{realSEeq}
\end{equation}
 The ion is at a fixed position and the subscript 0 indicates that there is no solute-solvent interaction in the statistical averaging. 
$U_{XS}$ is the solute-solvent interaction energy and is defined\cite{Ben-Naim1978,Ben-Amotz2005a}  as $U_{XS}=U_{X,N_s}-U_{N_s}$ where  $U_{X,N_s}$ is the total energy of the ion and solvent system including the electronic energy of the ion and $U_{N_s}$ gives the total energy of a given water structure with only the water molecules present. This expression is elaborated in section~A in the supplementary material (SM).

We can expand $U_{XS}$ for the case of a  charged hard sphere:
\begin{equation}
U_{XS}=U_{\text{Cav}}+U_{\text{PC}}
\end{equation}
where $U_{\text{Cav}}$ is a hard sphere interaction that excludes the oxygen atoms of the water molecules from some spherical region.
We can then write the free energy of solvation (see section~B of the SM) as: 
\begin{equation}
\begin{split}
\mu^*_{X}&=-k_{\text{B}}T\ln \left< e ^{-\beta U_{\text{\text{Cav}}}}\right>_{0}-k_{\text{B}}T\ln \left< e ^{-\beta U_{\text{PC}}}\right>_{U_{\text{\text{Cav}}}}\\&=\mu^*_{\text{Cav}}+\mu^*_{\text{PC}}
\label{SEpart}
\end{split}
\end{equation} 
We have dropped $E^\text{Vac}_X$ as it is zero for a point charge with no electrons. This partitioning is very useful as it simplifies the statistical treatment required. This is why it is a key piece of the quasi-chemical theory (QCT).\cite{Beck2006,Rogers2008}
We can estimate $\mu^*_{\text{Cav}}$ directly from simulation for cavities up to 3 - 4 $\text{ \AA}$ with a given water model by calculating the probability of cavity formation with an equilibrium simulation.\cite{Beck2006}
\begin{equation}
\mu^*_{\text{Cav}}=-k_{\text{B}}T\ln \left< e ^{-\beta U_{\text{\text{Cav}}}}\right>_{0}=-k_{\text{B}}T\ln p_0(R_{\text{Cav}})
\end{equation}
where $p_0(R_{\text{Cav}})$ is the probability of finding a cavity of size $R_{\text{Cav}}$ in bulk water. The energy of forming the cavity has been estimated  on the basis of classical-MD and  DFT-MD   calculations.\cite{Asthagiri2003a,Sabo2008}

\begin{figure}
        \centering
	     \begin{subfigure}{.5\textwidth}
                \includegraphics[width=.8\textwidth]{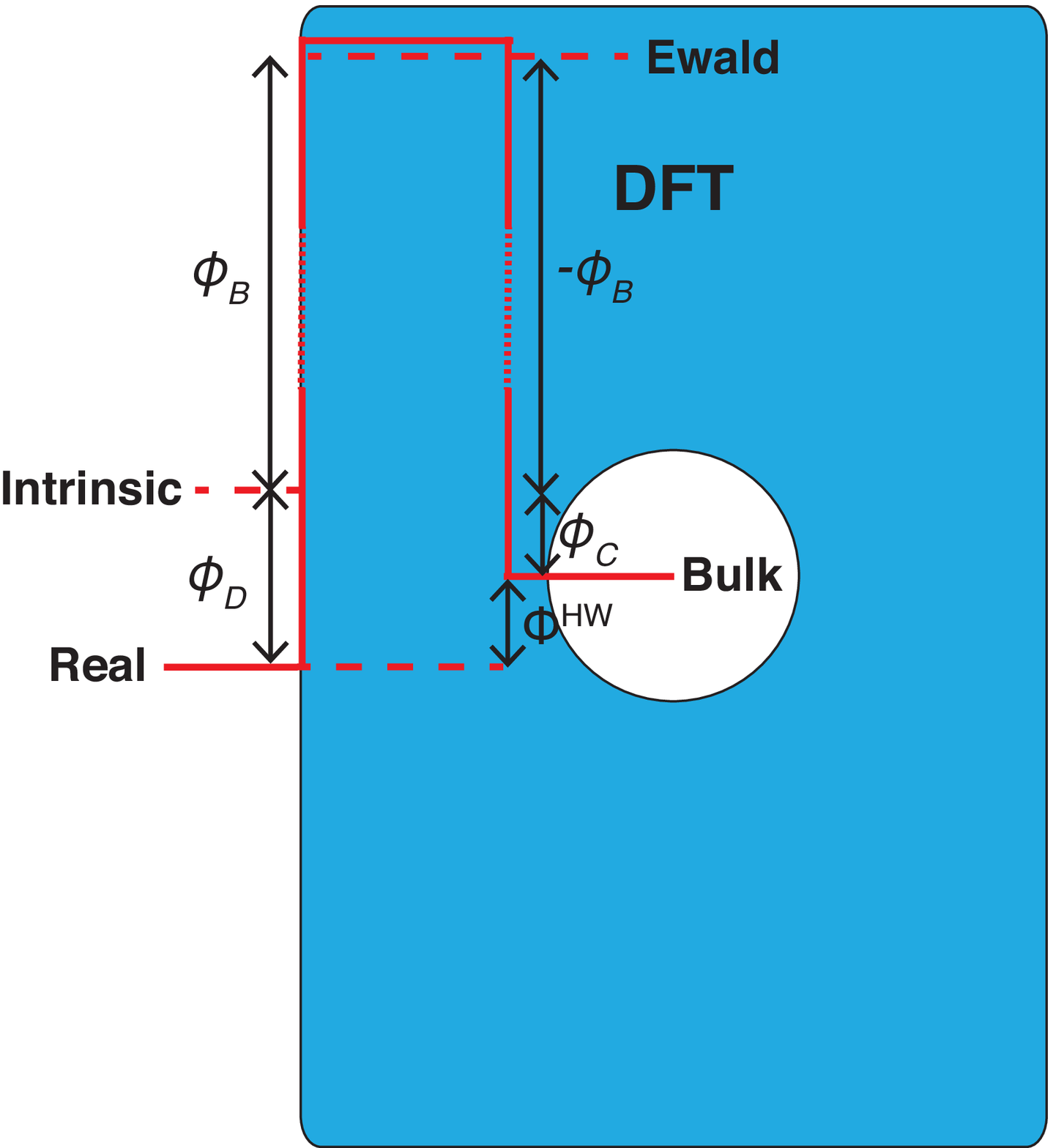}
        \end{subfigure}%
            \begin{subfigure}{.5\textwidth}
                \includegraphics[width=.8\textwidth]{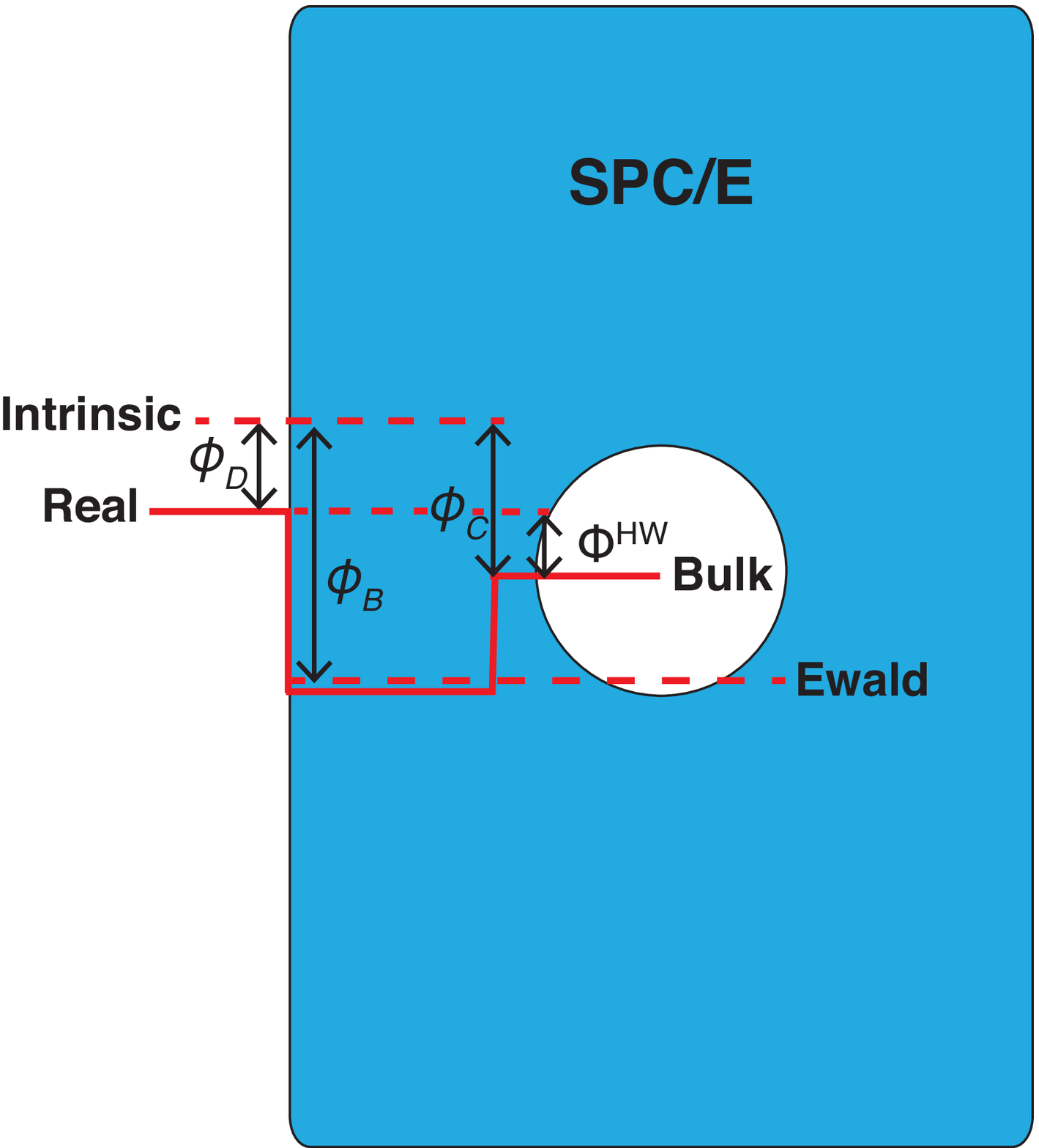}
        \end{subfigure}%
        \caption[]{A schematic of the contributions to the electrostatic potential for a cavity in water (not to scale) for both DFT and SPC/E water. The difference between the four different definitions of the solvation free energy is shown by labeling the corresponding zero of the potential for each. The Ewald values assume that the zero of the potential is slightly below the average potential inside bulk water. This is because the cavity lowers the average potential inside the cell slightly. }
        \label{surfacepotentialsfig}
\end{figure}

 $U_{\text{PC}}$ is the electrostatic interaction energy of the charge. It is given by\cite{Ben-Naim1978} $U_{\text{PC}}=U_{\text{PC},N_s}-U_{N_s}$, \emph{i.e.}, it is the difference in energy of a water structure with only waters present and with the waters and a point charge present and should only be evaluated when combined with a repulsive term. It is straightforward to calculate $\mu^*_{\text{PC}}$ by calculating the energy change as the charge is gradually turned on in increments of 0.1 or 0.05 $e$. The relevant expressions are provided in section~C of the SM. 
This expression assumes that the electrostatic potential is defined to be zero in the vapor phase infinitely far away from the air-water interface. We refer to this  as the `real' solvation free energy. ($\mu^*_{X}=\mu^{*\text{Real}}_{X}$) It corresponds to the actual (real) free energy change on taking an ion and moving it across the real air-water interface. 
The air-water interface creates a jump in the electrostatic potential. (See Figure~\ref{surfacepotentialsfig}) This is called the total surface potential.  In order to determine the zero of the electrostatic potential in the vapor phase relative to the aqueous phase it is important to have a reliable treatment of the air-water interface in order to properly estimate  this surface potential.  Any theoretical or experimental estimates of single ion solvation free energies that do not consider the air-water interface are not equivalent to the `real' solvation free energies as defined here.  There is currently no direct unambiguous experimental determination of this air-water surface potential. Different experiments give widely varying results and it is often unclear how the experimental measurement is related to the microscopic properties of the solution.\cite{Hunenberger2011}  

Recently several studies have examined the surface potential of water with DFT-MD.\cite{Leung2010,Kathmann2011,Sulpizi2013,Remsing2014} Particularly important is Ref.~\citenum{Remsing2014}, which gives the contributions to the potential inside a hard sphere cavity in water with revPBE-D3 and BLYP-D2. This study included the effect of the air-water interface by simulating a large water slab. This work makes it possible to  determine the contributions to the solvation free energy of single ions in water from the surface potential. A natural extension of  Ref.~\citenum{Remsing2014} is to place a point charge inside the hard sphere cavity and look at the response of the water to the presence of the charge. Using a water slab configuration for this calculation is problematic as the Coulomb interaction is long ranged and significantly perturbs the orientation of water molecules at the air-water interface in the finite simulation cell. This approach is therefore not useful to obtain the potentials inside a charged cavity and thus accurately determine solvation free energies. To circumvent this issue, it is necessary to perform the calculation of the charging free energies using periodic boundary conditions (PBC) under bulk solvation conditions, namely where no air-water interface is present. Ewald summation is an extremely useful method for treating electrostatics in PBC.  However, there are complications associated with any treatment of electrostatics in PBC such as correcting for finite size effects and determining the zero of the electrostatic potential. As a result, single ion solvation free energies calculated using Ewald summation must be carefully corrected before they can be considered physically meaningful. These corrections have been extensively outlined in the context of classical-MD studies\cite{Kastenholz2006,Reif2011a,Simonson2016} and it is relatively straightforward to apply these expressions to DFT-MD simulations. 

Decomposing the solvation free energy into one contribution from local water molecule interactions and another from the surface potential created by the air-water interface is  useful for providing  single ion solvation free  energies that can be used to parametrize simpler models of electrolyte solutions and for understanding ion-water interactions.  Unfortunately there is no clear accepted method for doing this in the literature. In fact there are at least three alternative definitions of the single ion solvation free energy with the surface potential removed that have been proposed. These different definitions correspond to different choices of the zero of the electrostatic potential. Differences in notation and nomenclature have made it a challenge to understand how the aforementioned different definitions are related. Here we aim to define and relate different approaches to computing single ion solvation free energies for direct and unambiguous comparison. 

\subsection{The Role of the Surface Potential in Single Ion Solvation free energies}

To understand the different definitions of solvation free energies, it is necessary to first understand the  contributions to the surface potential of a distant liquid-vapor interface. Comprehensive explanations of these contributions are available in the literature.\cite{Wilson1989,Pratt1992a,Hunenberger2011,Beck2013,Cendagorta2015} In brief, there is a dipolar surface potential, which is given by: 
\begin{equation}
\phi_\text{D}=-\epsilon_0^{-1}\int_{z_l}^{z_v} dz P_z(z)
\end{equation}
where $P_z(z)$ is the $z$ component of the average dipole density. \citeauthor{Hunenberger2011}\cite{Hunenberger2011} refer to this dipolar surface potential as $\chi_{M}$. It gives the change in the electrostatic potential caused by the average orientation and electronic polarization of water molecules at the air-water interface. 
The second contribution arises from the fact that there is a significant average potential inside a water molecule. Mathematically, it is given by the trace of the quadrupole moment of the water molecules.  We refer to this second contribution as the Bethe potential and refer to it with the symbol $\phi_\text{B}$. In System International (SI) units for a point charge model it is given by: 
\begin{equation}
\phi_\text{B}=-\frac{1}{6V\epsilon_0} \sum_i q \left<r^2\right>_i 
\label{BetheP}
\end{equation} 
or for a continuous charge distribution, $\rho (r)$,  it is given by:
\begin{equation}
\phi_\text{B}=-\frac{1}{6V\epsilon_0} \sum_i \int d^3 r  \rho (r)r^2
\label{ContBetheP}
\end{equation}
where V is the volume of the simulation box and the sum is over each atom in the box. The procedure for calculating this and the values for this quantity are given below. \citeauthor{Leung2010}\cite{Leung2010}  uses the notation $\phi_q$ to refer to this quantity, whereas \citeauthor{Hunenberger2011}\cite{Hunenberger2011} denote it as the exclusion potential, $-\xi$, although it has opposite sign.  It is also  referred to as the orientational disorder limit (ODL) correction ($-\Phi_{\text{ODL}}$).\cite{Kastenholz2006,Simonson2017}  The Bethe potential can be calculated solely from a bulk simulation of water, whereas the dipolar surface potential relies on an accurate simulation of the air-water interface. 

The sum of these two terms gives the total surface potential of the air-water interface:
\begin{equation}
\Delta\phi=\phi_\text{B}+\phi_\text{D}=-\epsilon_0^{-1}\int_{z_l}^{z_v} dz \rho(z)z
\end{equation}
where $\rho(z)$ is the average charge distribution as a function of position across the interface.
For \emph{ab initio} water treated with all electrons and no pseudo-potentials, this is a real physical quantity of around 3 to 4 V. It is much larger than most experimental estimates of the surface potential because almost all experiments do not probe the internal potential of the water molecules. Electron holography is a notable exception where a large potential of this size is confirmed.\cite{Kathmann2011}  However, for most classical water models this total surface potential is entirely unphysical as the average electrostatic potential inside a point charge model of water is opposite in sign compared with the average electrostatic potential inside real water. This is due to the large negative point charges that classical water models generally use. The total surface potential is often the only one reported in calculations of the surface potential.\cite{Lamoureux2006,Rick2016} Considering that the two contributions to the total surface potentials have very different physical origins it would provide more insight to provide them separately. 

This partitioning into a dipolar and Bethe potential depends on the choice of the origin of the water molecule.  This origin dependence is given by 
\begin{equation} 
\phi_\text{D}^\text{shift}=-\frac{\rho_w\left<\boldsymbol{\mu}\cdot\boldsymbol{d}\right>}{3\epsilon_0} 
\label{phishiftEq}
\end{equation}
where $\boldsymbol{d}$ gives the change in the origin position. This expression is derived in section~D of the SM and is generally applicable to any non-rigid solvent molecule including large flexible polymer molecules. This highlights the limitation of a center dependent description of ion solvation. In contrast to the dipolar and Bethe potential, the total surface potential does not depend on the choice of the origin. 

An alternative definition of the solvation free energy that does not require an accurate treatment of the air-water interface exists. \citeauthor{Hunenberger2011}\cite{Hunenberger2011} refer  to it as the \textit{intrinsic} solvation free energy and it corresponds to the solvation free energy that results if you  subtract  the dipolar surface potential from the   `real' solvation free energies:
\begin{equation}
\mu_{X}^{*\text{Int}}=\mu^*_{X}-q_I\phi_\text{D}
\label{IntSE}
\end{equation} 
More intuitively,  the intrinsic solvation free energy corresponds to choosing the zero of the electrostatic potential to be in the vapor phase infinitely far away from the air-water surface assuming that the molecules at the interface are isotropically oriented, \emph{i.e.}, $\phi_\text{D}=0$. The problem with this quantity is that the notion of `isotropically oriented' depends on the choice of the molecular center for the same reason that the dipolar surface potential depends on the choice of the origin of the water molecule. Hence, the intrinsic solvation free energies do as well. It has been shown\cite{Hunenberger2011} that using  a molecular (M-type) based cut-off in the summation of the Coulomb potential results in this definition of the intrinsic solvation free energy. Using a molecular based cut-off effectively creates an air-water surface with zero dipolar surface potential (for the chosen water center). It has been shown that using an M-type cut off also results in solvation free energies that depend on the choice of the water molecule's origin.\cite{Hummer1998a,Aqvist1998}   

Generally, surface potential calculations, including Ref~\citenum{Remsing2014}, choose the oxygen atom to be  the center of the water molecule. With oxygen for the molecular center a value of 0.48 V is determined for the dipolar surface potential based on DFT-MD calculations. \cite{Remsing2014}  The standard choice of the oxygen atom as the molecular center is  arbitrary and chosen primarily for computational convenience. As such the resulting intrinsic solvation free energies are unlikely to carry any real physical significance and will only be useful for comparison with other computational methods. It is not clear that there is any means of  determining what the best choice for the molecular center is. A potentially physically meaningful choice is to place the origin at the center of nuclear charge, which is the analog of the center of mass but with the mass replaced by the nuclear charge. Using this choice, Eq.~\ref{phishiftEq} gives an increase in the Bethe potential of 0.14 V and a corresponding decrease in the dipolar surface potential of 0.14 V. As a result the intrinsic solvation free energies of cations become less negative, whereas the real and bulk quantities stay the same.


Some researchers simply use Ewald summation to calculate the ionic solvation free energies without correcting for the surface potentials or finite size effects. Often these values are referred to as the intrinsic solvation free energies.\cite{Lamoureux2006} However, it is useful to have a separate term for these values in order to distinguish them from the intrinsic solvation free energies defined above. We therefore refer to them as Ewald solvation free energies. They are related to the `real' solvation free energies by the following expression:
\begin{equation}
\mu_{X}^{\text{*Ewald}}=\mu^*_{X}-q_I\phi_\text{D}-q_I \phi_\text{B}+\mu_\text{Ew-Corr}
\label{EwSE}
\end{equation} 
The Ewald values are the most commonly reported values in molecular simulation studies.\cite{Grossfield2003,Rajamani2004,Lamoureux2006,Bardhan2012,Sedlmeier2013} Ewald summation sets the zero of the electrostatic potential to be the average  potential over one unit cell. \cite{Kastenholz2006} This is an arbitrary definition of the zero of the electrostatic potential as it depends on the internal structure of the water molecules and varies dramatically between different representations of water.  For example, there is a 4 V difference between quantum and classical water models. The details for the calculation of the Bethe potential and the values for the systems studied here are given in section~E of the SM. If an atom-based cutoff in the summation of the Coulomb interaction is used (P-type summation), then the resulting solvation free energies are equivalent to the Ewald solvation free energies.\cite{Reif2011a}
When the raw energies calculated with Ewald summation are used in Eq.~\ref{realSEeq} to calculate the solvation free energies then there is an additional correction ($\mu_\text{Ew-Corr}$) described in section~F of the SM.
\citeauthor{Reif2016}\cite{Hunenberger2011,Reif2016} have provided extensive justification for why the Ewald solvation free energies are not a useful concept. The most significant reason is that they inherently include a large and arbitrary contribution associated with the internal properties of the water molecule. 

The final type of single ion solvation free energy was first defined by Beck.\cite{Beck2013} They are called the \textit{bulk} solvation free energies and  assume that the electrostatic potential is zero at a point at the center of an uncharged cavity carved out of water. 
To define this quantity we must introduce $\phi_\text{C}$, which is the average potential created inside a cavity due to the average orientation of water molecules around that cavity. It is defined mathematically in the SI of Ref.~\citenum{Remsing2014} in terms of a traceless multipole moment expansion. \citeauthor{Hunenberger2011}\cite{Hunenberger2011} refer to this as $\zeta_{M}$
This potential is essentially the difference between the electrostatic potential in a cavity and the potential in bulk water minus the Bethe potential. As the cavity approaches macroscopic size $\phi_\text{C}$ must converge to $-\phi_\text{D}$. As it is defined in terms of traceless multipole moments, $\phi_\text{C}$ does not depend on the Bethe potential. 

This then allows for the definition of the net potential, which Beck refers to as:   $\phi_{np}$. Ref.~\citenum{Remsing2014} refers to it as:  $\Phi^{\text{HW}}$. It is given by:  $\Phi^{\text{HW}}=\phi_\text{C}+\phi_\text{D}$ or equivalently $\phi_{np}=\phi_{lp}+\phi_{sp}$ in Beck's notation. (Although note that  $\phi_{lp}$ and $\phi_{sp}$ include compensating contributions from the Bethe potential, \emph{i.e.},  $\phi_{lp}=\phi_\text{C}+\phi_\text{B}$) It corresponds to the difference in potential between the vapor phase and a small cavity formed in water. The bulk solvation free energies are then defined as:
\begin{equation}
\mu_{X}^{*\text{Bulk}}=\mu^*_{X}-q_I\Phi^{\text{HW}}
\label{BulkSE}
\end{equation} 
This net potential is inherently dependent on the size of the ion and the nature of how it repels the water molecules, as such it does not have a single value.\cite{Remsing2014,Ashbaugh2000a}

\subsection{The Connection to Born Theory}
Eq.~\ref{BulkSE} is useful because various approaches to determining the solvation free energy do not include a contribution from any surface potentials. The assumption that the electrostatic potential is zero at the center of an uncharged cavity in water implies that the solvation free energy of a charge at the center of that cavity is purely quadratic in the charge, \emph{i.e.}, it has no linear contribution, namely $\left.\frac{d\mu_{X}^{*\text{Bulk}}}{dq}\right|_{q\to0}=0$. Two important examples are the Born model and the TATB assumption that both implicitly assume a net potential of zero. Following Beck\cite{Beck2013}, it is appropriate to compare Born model calculations with the bulk solvation free energies not the intrinsic values.  
Bulk solvation free energies are only useful if the charging process follows linear response  as only then is there any point splitting the solvation free energy into a linear term and a quadratic term with respect to the charge.
Tables~\ref{SPDef}, \ref{SEDef}, \ref{SPNot}, and Figure~\ref{surfacepotentialsfig}  summarize the information provided in this section. 
\begin{table}
\centering
\caption[]{Surface potential definitions}
 \begin{tabular*}{1\textwidth}{@{\extracolsep{\fill}}clclc}
\hline
Type &Expression\\ \hline 
Dipolar&$ \phi_\text{D}$=$-\epsilon_0^{-1}\int_{z_l}^{z_v} dz P_z(z)$ \\
Bethe&$\phi_\text{B}$=$-\frac{1}{6V\epsilon_0} \sum_i q \left<r^2\right>_i $\\
Cavity&$ \phi_\text{C}$=See SI of Ref.\citenum{Remsing2014}\\
Net&$ \Phi^{\text{HW}}$=$\phi_\text{C}+\phi_\text{D}$\\
Total&$ \Delta\phi$=$\phi_\text{D}+\phi_\text{B}=-\epsilon_0^{-1}\int_{z_l}^{z_v} dz \rho(z)z$\\
\hline
\end{tabular*}
\label{SPDef}
\end{table}

\begin{table}
\centering
\caption[]{Four types of solvation free energies}
 \begin{tabular*}{1\textwidth}{@{\extracolsep{\fill}}clclclclc}
\hline
Type&Expression \\ \hline 
Real &$\mu^*_{X}$ \\
Intrinsic &$\mu^{*\text{Int}}_{X}$  =$\mu^*_{X}-q_I\phi_\text{D}$ \\
Bulk &$\mu_{X}^{*\text{Bulk}}$ =$\mu^*_{X}-q_I\Phi^{\text{HW}}$\\
Ewald & $\mu_{X}^{*\text{Ewald}}$=$ \mu^*_{X}-q_I\phi_\text{D}-q_I \phi_\text{B} +\mu_\text{Ew-Corr}$ \\ \hline
\end{tabular*}
\label{SEDef}
\end{table}

\begin{table}
\centering
\caption[]{Surface potential notations}
 \begin{tabular*}{1\textwidth}{@{\extracolsep{\fill}}clclclc}
\hline
\citeauthor{Remsing2014}\cite{Remsing2014} &\citeauthor{Hunenberger2011}\cite{Hunenberger2011}&\citeauthor{Beck2013}\cite{Beck2013}&\citeauthor{Simonson2017}\cite{Simonson2017}\\ \hline 
$ \phi_\text{D}$&$\chi_{M}$&-&-\\
$\phi_\text{B}$ &-$\xi$ (Exclusion Potential)&-&$-\Phi_{\text{PBC}}^{\text{M-sum}}$ or $-\Phi_{\text{ODL}}$ \\
$\Delta \phi =\phi_\text{D}+\phi_\text{B}$&$\chi_{P}$&$\phi_{sp}$&$\Phi_{\text{lv}}$ or $\Phi_{G}$ (Galvani)\\
$ \phi_\text{C}+\phi_\text{B}$&$\zeta_{P}$&$\phi_{lp}$&-\\
$ \phi_\text{C}$&$\zeta_{M}$&-&-\\
$ \Phi^{\text{HW}}$&-&$\phi_{np}$&-\\
\hline
\end{tabular*}
\label{SPNot}
\end{table}

In order to clarify the connection to Born theory it is  useful to define an effective potential:
\begin{equation}
\phi_\text{eff}(q)=\frac{d\mu^*_{X}}{dq}= \left<\phi_0+\frac{\phi_I(q)}{2}+\frac{q}{2}\frac{d\phi_I(q)}{d q} \right>_{U_{\text{Cav}}+U_\text{q}}
\label{phieff}
\end{equation}
This is derived in section~G of the SM. This expression allows us to use a Taylor expansion to write the solvation free energy as:
\begin{equation}
\mu^*_{\text{PC}}=q \phi_\text{eff}(0) +\left.\frac{q^2}{2}\frac{d\phi_\text{eff}}{dq} \right|_{q\to0}+O(q^3)
\label{SETaylor}
\end{equation}
The second two terms in the brackets in Eq.~\ref{phieff} are proportional to the charge and will go to zero as the charge goes to zero. Hence, we can see that $ \phi_\text{eff}(0) =\left<\phi_0\right>_{U_{\text{Cav}}}$. This is just the net potential ($\Phi^\text{HW}$) assuming we are calculating `real' solvation free energies. It follows that the bulk free energies are given by:
\begin{equation}
\mu^{*\text{Bulk}}_{\text{PC}}=\mu^*_{\text{PC}}-q\Phi^\text{HW}=\left.\frac{q^2}{2}\frac{d\phi_\text{eff}}{dq} \right|_{q\to0}+O(q^3)
\label{SETaylorBulk}
\end{equation}
Hence, the bulk solvation free energies correspond to the solvation free energies with the linear term removed. 
The Born equation is:
\begin{equation}
\mu_{\text{Born}}=-\frac{q^2}{8\pi\epsilon_oR_{\text{Born}}}\left(1-\frac{1}{\epsilon_r}\right)
\label{BornSE}
\end{equation}
Comparing this with Eq.~\ref{SETaylorBulk} shows that the Born model should be compared with the bulk solvation free energies. We can equate these two expressions to derive an expression for the Born radius that can be determined directly from simulation:
\begin{equation}
R_{\text{Born}}=-\frac{1}{4\pi\epsilon_o}\left(1-\frac{1}{\epsilon_r}\right)\left(\frac{d\phi_\text{eff}}{dq}\right)^{-1}
\label{BornRad}
\end{equation}

A valuable extension, not carried out here, would be to partition $U_{\text{PC}}$ up into long and short-range contributions  using Local Molecular Field theory.\cite{Remsing2011} Ref.~\citenum{beck2011} and Ref.~\citenum{beck2011a} are two examples where this partitioning is performed for classical water models to provide physical insight. A new method of carrying out this partitioning has recently been put forward by Remsing and Weeks.\cite{Remsing2016} This method begins by solvating a diffuse smooth Gaussian charge density in bulk water. The advantage of this approach is that the solvation of the Gaussian is a linear process that can be estimated analytically using dielectric continuum theory.  However, to finish the process, the Gaussian must be collapsed down to a point charge, which is a complex non-linear process and so is difficult to evaluate with simulation. In contrast, as shown below, simply turning a charge on in a cavity shows only small non-linearities. 

\subsection{Caveats}
The single ion solvation free energies calculated here assume that the ions are in the insulating phase.  The complexity of properly treating the electrolyte solution as a conductor has been discussed in Ref.~\citenum{you2014}.  Because conductors are subject to the electro-neutrality condition, two electrolyte solutions in equilibrium can produce a potential of the phase that is fundamentally different from the surface potentials discussed here.\cite{you2014,Levin2008a} There is disagreement about what the the potential of the phase goes to in the limit of infinite dilution. Ref.~\citenum{Levin2008a} uses the canonical distribution which is appropriate for real finite Coulomb systems\cite{Bobrov2012} and shows that the potential of the phase goes to zero in the limit of infinite dilution. Ref.~\citenum{you2014} takes a different position but does not explicitly derive the potential of the phase by minimizing the Helmholtz free energy of a finite system as Ref.~\citenum{Levin2008a} does.

A reviewer has raised the concern that the single ion solvation free defined here and elsewhere are in violation of the Gibbs-Guggenheim principle (GGP),\cite{pethica2007} which states that the electrical potential difference between two regions of different chemical composition cannot be measured. Ref.~\citenum{pethica2007} states that real single ion solvation free energies are obtainable from experimental measurements subject to certain reasonable assumptions which do not violate the GGP such as that `single ion activity coefficients are equal to the (measurable) mean ionic activity coefficients for the electrolyte.' The experimental accessibility of the real single ion solvation free energies is also discussed extensively in section 4.5.2 of Ref.~\citenum{Hunenberger2011}.

The real single ion solvation free energies are therefore a physically meaningful and measurable quantity. The intrinsic and bulk solvation free energies can be defined explicitly in statistical mechanical terms, but in accordance with the GGP it is not clear that they correspond to any physically measurable process as they are defined in terms of surface potentials. These values are still useful conceptually as they provide a means of comparing different theoretical methods at an equivalent level. For example, intrinsic solvation free energies calculated with DFT-MD and classical-MD can be compared assuming the same molecular origin is used. An additional caveat is that the `real' solvation free energy as defined here cannot be straightforwardly generalized to situations where the air-water interface is unstable such as water above the critical point.

\section{Calculation details}
The system contained 96 water molecules and a hard sphere with a charge in the center of a 14.3$^3$~\AA$^{3}$  supercell. 
To determine the box size we used the expression: 
\begin{equation}
L=\left(\frac{N_w}{\rho_w}+\frac{4\pi}{3}R_I^3\right)^{-3}
\end{equation}
which gives 14.3~\AA\  for both a 2 \AA\ and a 2.6 \AA\ cavity.
NPT simulations were run with the uncharged 2  \AA\ cavity present to test this choice of the box size. The revPBE-D3 simulations had an average of 14.3 \AA\  agreeing with this estimate. The BLYP-D2 simulations were slightly lower at 14.0.  \AA\   The revPBE-D3 functional is believed to give a better estimate of the density of bulk water, and so 14.3  \AA\  was used for all the NVT calculations so that the effect of the change in the density with the functional was not included. 

To model a charged hard sphere in \texttt{CP2K}  we used a hydrogen atom with its core charge scaled to the desired value. No basis functions are placed on the hydrogen atom as  otherwise electrons will transfer to it.

The hydrogen atom sits at the center of a hard sphere repulsive interaction that acts only on the oxygen atoms and is given, in a.u., by:
\begin{equation}
U_{\text{Cav}}=\sum_{O}1-\tanh\left(\left(r_{\text{XO}}-R_\text{Cav}\right)/0.05\right)
\end{equation}
where $r_{\text{XO}}$ is the ion oxygen distance and $R_\text{Cav}$ is the hard sphere (cavity) radius. 

To calculate the Bethe potentials we chose the oxygen atom to be the center of the water molecule for consistency with Ref.~\citenum{Remsing2014}.  Eq.~\ref{BetheP} was used to calculate the contribution from the hydrogen atoms and from the electrons where the positions of the electron pairs were taken to be the location of the Wannier Centers. The Wannier spreads\cite{Berghold2000,Remsing2014} were then added to account for the finite spread of the electron pairs. The contribution from the spatial spread of the pseudo-potentials was estimated using Eq.~\ref{ContBetheP}.

The Ewald solvation free energies were calculated using the raw energy differences output from \texttt{CP2K}. The energy differences were used in Eq.~27 and Eq.~28 of the SM and were then corrected using Eq.~\ref{EwSE} to calculate the `real' solvation free energy. Eq.~\ref{IntSE} and Eq.~\ref{BulkSE} were then used to calculate the intrinsic and bulk solvation free energies respectively.   The Ewald correction requires a value for the size of the ion. Based on the recommendation in Ref.~\citenum{Reif2011}, this is given best by the mean of the peak position in the ion-oxygen radial distribution function (RDF) and the Goldschmidt in-crystal radii. This results in values of  1.41, 2.01 and 2.09 for lithium, fluoride and potassium respectively, where we have used the experimental peak position and crystal radii given in Ref.~\citenum{Duignan2013a}.

We can estimate the net potentials using the Hartree potential calculated in \texttt{CP2K}. We take the value at the center of the cavity and then add the Bethe and dipolar potentials to arrive at the real net potential properly referenced to the vapor phase. 

The NVT simulations (at 300~K) were performed under PBC using  the \texttt{CP2K} simulation suite (http:www.cp2k.org) with the \emph{QuickStep} module for the DFT calculations.~\cite{VandeVondele2005} Shorter range double zeta basis sets optimized for the condensed phase\cite{VandeVondele2007} were used in conjunction with Goedecker-Teter-Hutter (GTH) pseudopotentials~\cite{Goedecker1996} and a 400~Ry cutoff for the auxiliary plane wave basis.  A Nos\'e--Hoover thermostat was attached to every degree of freedom to ensure equilibration.\cite{Martyna1992} Two different DFT functionals were used, one was the Becke exchange and correlation due to Lee, Yang, and Parr (BLYP)\cite{Becke1988,Lee1988} and the other was the revised Perdew, Burke, and Ernzerhof (revPBE).\cite{Perdew1996,Zhang1998}  The D2 and D3 dispersion corrections due to Grimme\cite{Grimme2004,Grimme2010} were used for BLYP and revPBE respectively. A 0.5 fs time step was used.  The energies were accumulated for $\approx$ 12~ps after $\approx$ 3~ps of equilibration for each charge increment. We use the standard Ewald summation method to treat the electrostatics as implemented in CP2K and described in Ref.~\citenum{VandeVondele2005} and in the CP2K manual. No alternatives were considered or tested.

The determination of the error is very challenging due to the highly correlated nature of the data combined with the short trajectories used and the fact that the solvation free energies depend on the fluctuations of the energy not just on the average and so  blocking or Monte Carlo bootstrapping methodologies are ineffective. We therefore use a more heuristic approach and simply take the difference in the energy of charging vs. decharging the ions as an estimate of the uncertainty for each step of the charging process. The propagation of these errors  provides the estimate for the uncertainty in the total Ewald solvation free energies given in Table~\ref{SETable}.
The other sources of error that were considered were first how the solvation free energy of reasonably sized subsets of the data varies, second how the data converges as the length of the trajectory is increased, and third  what the effect of increasing the equilibration time is. These errors were all smaller than the uncertainty determined using the difference between charging and decharging. The probability distributions of the energy and the potential at the center of the uncharged cavity for some representative cases are shown in section~H of the SM indicating that any non Gaussian behavior is relatively small and should not effect the results. Ref.~\citenum{Shi2013} has shown that higher order cumulants of the cavity potential fluctuations can make significant contributions to the ionic solvation free energies. There is an additional source of error associated with the choice of the radius of the ion in the expression for the finite size correction (Eq.~37 of the SM). \citeauthor{Reif2011}\cite{Reif2011} claim that this error is no larger than 1 kJmol$^{-1}$ The uncertainties do not account for the uncertainty associated with the physical approximations made to perform the calculations such as the Born-Oppenheimer approximation with classical motion for the nuclei or the use of generalized gradient corrected functionals with pseudo-potentials for the electronic energy.  Ref.~\citenum{Wilkins2015} indicates that nuclear quantum effects are reasonably small ($\approx 4$ kJ mol$^{-1}$) for these systems although confirmation with path integral DFT-MD should be performed. 

\section{Results and Discussion}
The cavity formation energy can be estimated straightforwardly by looking at the cavity formation probability in bulk water. This has been done before for DFT-MD simulation.\cite{Asthagiri2003a} For the small cavities studied here, classical-MD\cite{Sabo2008} appears to give a reasonable estimate of this contribution.  Because the focus of this work is the charging free energies and this term is independent of the charge, we do not provide an estimate of it here.

The values for the net potential ($\phi_\text{HW}$) are shown in Table~\ref{HWPots} and are semi-quantitatively consistent with the estimates from Ref.~\citenum{Remsing2014} where the same quantity was calculated using a water slab.
\begin{table}
\begin{threeparttable}
\centering
\caption[]{Net Potentials ($\phi_\text{HW}$) calculated using the Hartree potential corrected with the Bethe and dipolar surface potential  }
 \begin{tabular*}{1\textwidth}{@{\extracolsep{\fill}}clclclclc}\hline
 Functional &Cavity size(\AA)&$\Phi_\text{HW}(V)$ \\ \hline
revPBE-D3& 2.0 &0.19\\
revPBE-D3& 2.85 (He)\footnote{Calculated using a helium atom to create the cavity rather than using a hard sphere repulsion} &0.14 \\
revPBE-D3& 2.6 &0.04 \\
BLYP-D2& 2.0&-0.03  \\
\hline
\end{tabular*}
\label{HWPots}
\end{threeparttable}
\end{table}
The estimates here are slightly lower by about 0.1 V than in Ref.~\citenum{Remsing2014}. 
We can therefore confidently state that the net potential of DFT water is small ($\approx 0$ V $-$ 0.2  V). One drawback of the electrostatic potentials calculated in Ref.~\citenum{Remsing2014} is that a hard sphere repulsion that acts only on the oxygen atoms may be unphysical because a real solute will repel the electron density and so the orientation of the water molecules around the real solute could be different.\cite{Pollard2014a}  We have addressed this critique by examining the potential created by water surrounding a helium atom. The resulting potential is very similar, indicating that this detail does not significantly alter the cavity electrostatics. An uncertainty of approximately 0.1 V for the values of the net potential can be estimated based on the agreement of the different methods of calculating these values.

We have calculated the solvation free energy of a 2 \AA\ cation and a 2.6 \AA\ cation and anion. These were chosen as they are close in size to the lithium, potassium and fluoride ions respectively.  The solvation free energies of neutral pairs of these ions and the differences in solvation free energies of ions of the same charge are  given in Table~\ref{SETablediff}. These values are approximately independent of the choice of solvation free energy type because the surface potential terms cancel out. The reason for this is clear from Table~\ref{SEDef}. These values are also directly and unambiguously experimentally accessible. 
\begin{table}
\begin{threeparttable}
\centering
\caption[]{Electrostatic solvation free energies differences and sums calculated with revPBE-D3. Values are given in kJmol$^{-1}$. }
 \begin{tabular*}{1\textwidth}{@{\extracolsep{\fill}}lclclclclclc}\hline
&$\mu^{*\text{Real}}_{\text{PC}}$&$\mu_{\text{PC}}^{*\text{Bulk}}$&$\mu_{\text{PC}}^{*\text{Int.}}$&$\mu_{\text{PC}}^{*\text{Ewald}}$\\ \hline
$\mu^{*}_{2.0^+}-\mu^{*}_{2.6^+}$&$-156\pm4$&$-171\pm3$&$-156\pm3$&$-154\pm3$ \\
$\mu^{*}_{2.0^+}+\mu^{*}_{2.6^-}$&$-1105 \pm 4$&$-1120 \pm 4 $&$-1105 \pm 4$&$-1115 \pm 4 $\\
$\mu^{*}_{2.6^+}+\mu^{*}_{2.6^-}$ &$-949 \pm 2 $&$-949\pm2$&$-949 \pm 2 $&$-962 \pm 4$\\
\hline
\end{tabular*}
\label{SETablediff}
\end{threeparttable}
\end{table}

\begin{figure}
        \centering
	\begin{subfigure}{0.5\textwidth}
                \includegraphics[clip,width=\textwidth]{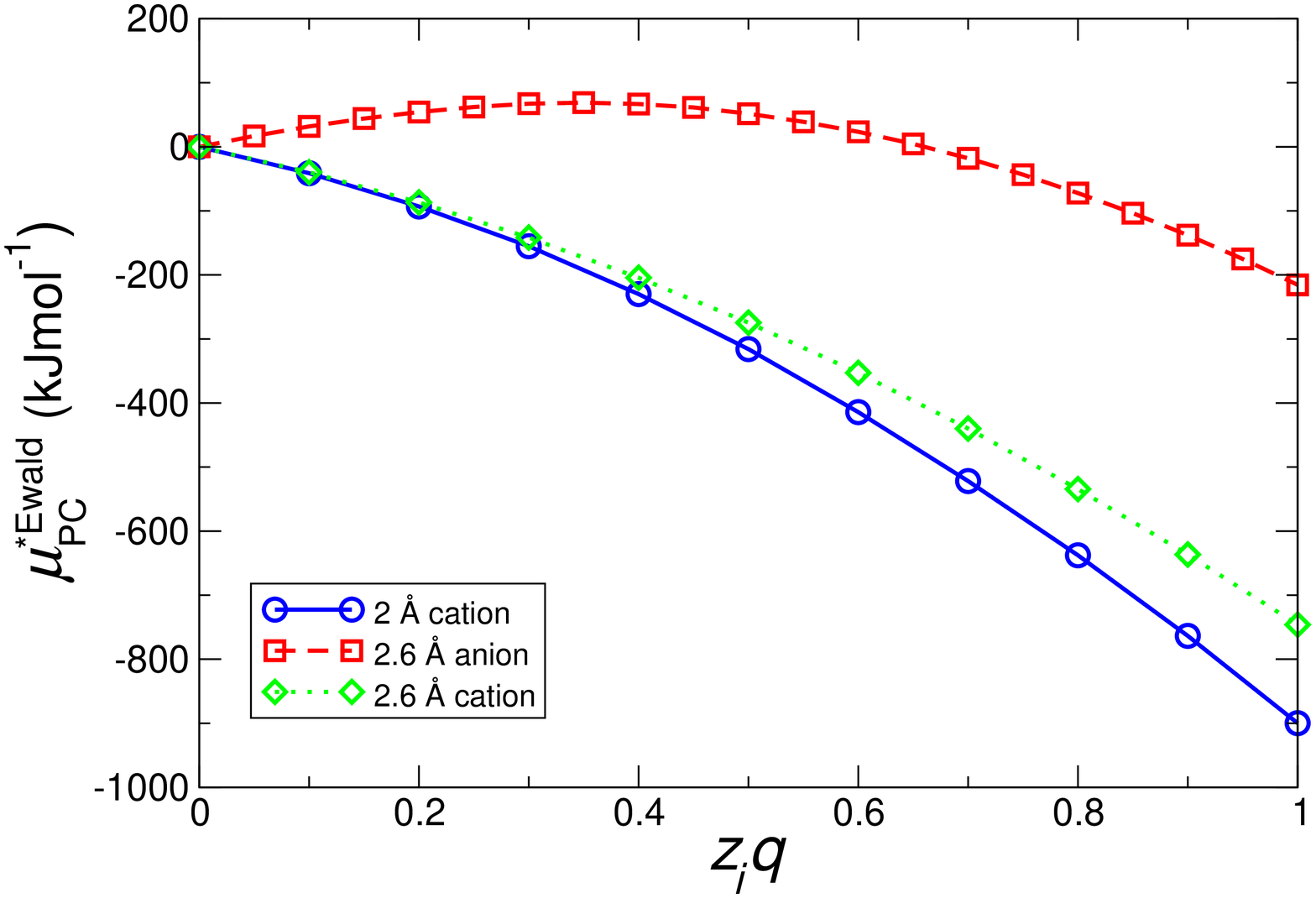}
                \caption{Ewald solvation free energy}
                \label{EwSEfig}
        \end{subfigure}%
	~
	     \begin{subfigure}{0.5\textwidth}
                \includegraphics[clip,width=\textwidth]{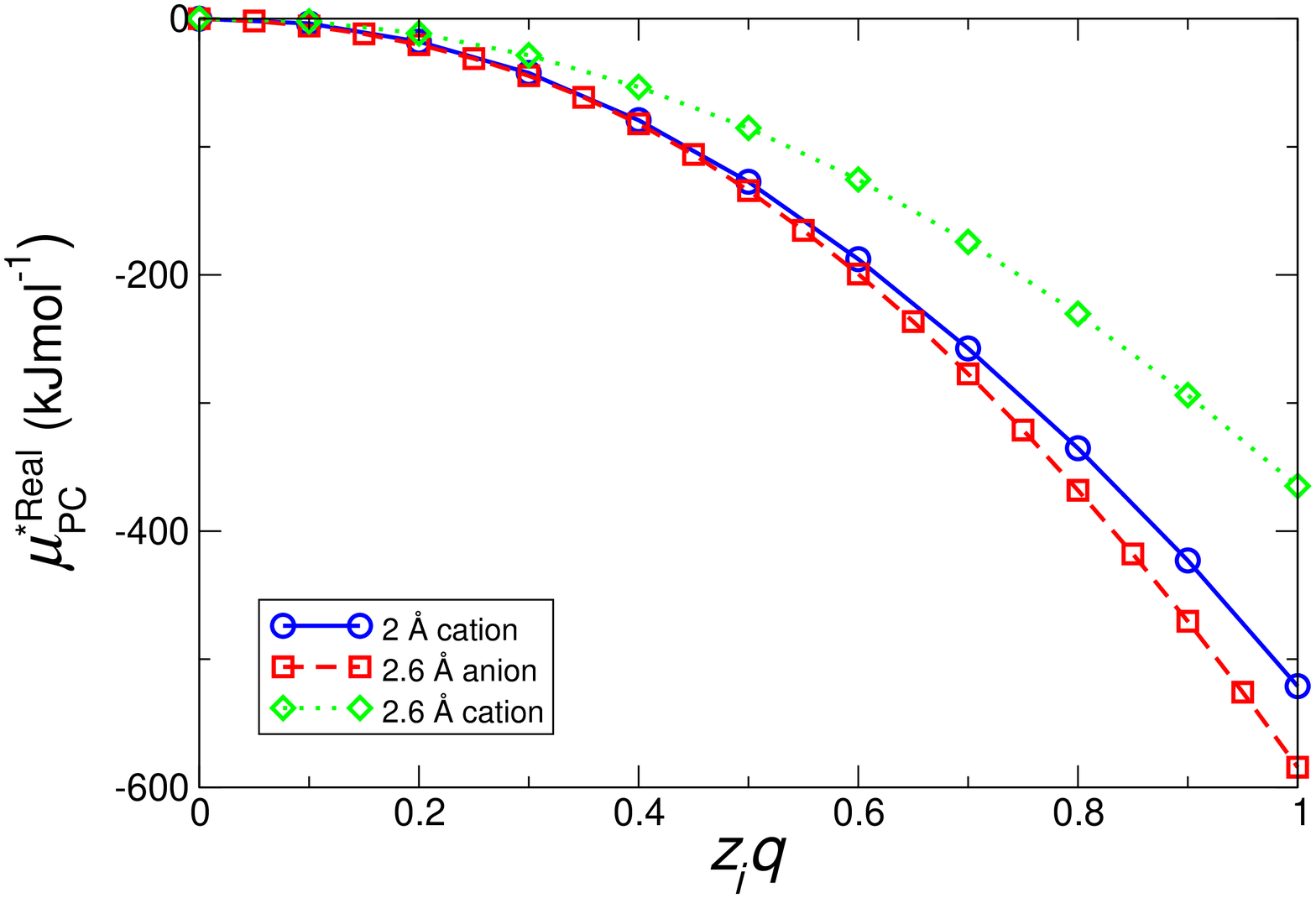}
                \caption{`Real' solvation free energy}
                \label{RealSEfig}
        \end{subfigure}%
        \caption[]{Solvation free energies as a function of charge for a 2 \AA\ cation and a 2.6 \AA\ cation and anion. }
        \label{SEfig}
\end{figure}

\begin{table}
\begin{threeparttable}
\centering
\caption[]{Electrostatic solvation free energies calculated with revPBE-D3. Values are given in kJmol$^{-1}$. }
 \begin{tabular*}{1\textwidth}{@{\extracolsep{\fill}}lclclclclclc}\hline
Charge&Cavity size(\AA)&$\mu^{*\text{Real}}_{\text{PC}}$&$\mu_{\text{PC}}^{*\text{Bulk}}$&$\mu_{\text{PC}}^{*\text{Int.}}$&$\mu_{\text{PC}}^{*\text{Ewald}}$\\ \hline
 \ \ \ \ $+$&2.0 &$-521\pm3$&$-539\pm3$&$-567\pm3$&$-900\pm3$ \\
 \ \ \ \ $+$&2.6 & $-365 \pm 1 $&$-369 \pm 1 $&$-411 \pm 1 $&$-746 \pm 1 $\\
 \ \ \ \ $-$&2.6 & $-584 \pm 2$&$-580\pm2$&$-538 \pm 2$&$-215 \pm 2$\\
\hline
\end{tabular*}
\label{SETable}
\end{threeparttable}
\end{table}
Figure~\ref{SEfig} and Table~\ref{SETable} give the single ion solvation free energies of these ions. It is clear that the single ion solvation free energies calculated with Ewald summation are unphysical. They are much too large for the cations and much too small for the anion.
This is due to the very large Bethe potential of \textit{ab initio} water and it highlights that the single ion Ewald solvation free energies do not correspond to an experimentally measurable property. For classical water models the Bethe potential is substantially smaller and so these values seem much more reasonable. Many researchers have  reported these values without making it clear that they do not correspond to an experimentally measurable property.\cite{Grossfield2003,Rajamani2004,Lamoureux2006,Bardhan2012,Sedlmeier2013}  As \citeauthor{Hunenberger2011}\cite{Hunenberger2011} argue, this has lead to confusion in the literature.\cite{Hunenberger2011} The single ion Ewald solvation free energies are not relevant to experiment or even theoretical comparison as they can only be compared in cases where the same methodology and water model have been used. It is better to report the `real', intrinsic and bulk solvation free energies as provided in Table~\ref{SETable}. 

\begin{figure}
        \centering
                \includegraphics[clip,width=.66\textwidth]{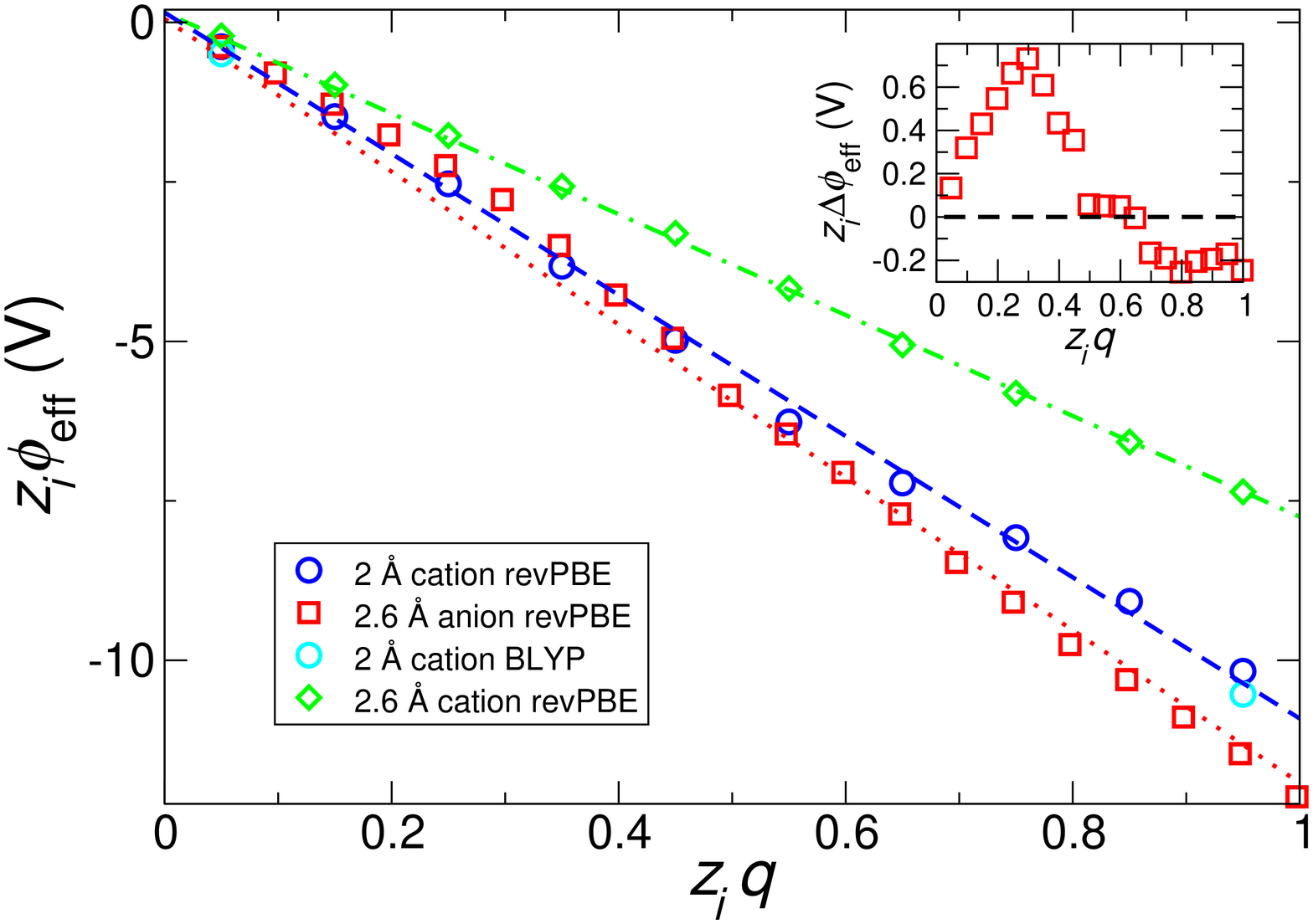}
        \caption[]{Effective potential (Eq.~\ref{phieff}) for 2 \AA\ cation and  2.6 \AA\ cation and anion as a function of charge. Inset shows the difference between the 2.6 \AA\  anion and the linear model highlighting the non-linearity at low charge. }
        \label{derivRealSE}
\end{figure}

Figure~\ref{derivRealSE} shows that the effective potential appears to be approximately linear for cations as a function of charge. This is consistent with classical-MD.\cite{Bardhan2012} This linearity will likely break down for multivalent ions due to dielectric saturation.  The inset of Figure~\ref{derivRealSE} shows that there does appear to be some non-linearity in the low charge region for the 2.6 \AA\ anion, which is also consistent with some classical-MD studies.\cite{Hummer1998b}

\begin{table}
\begin{threeparttable}
\centering
\caption[]{Linear models of the solvation free energy}
 \begin{tabular*}{1\textwidth}{@{\extracolsep{\fill}}lclclclclclclc}\hline
Charge&Cavity size(\AA)&$z_i\phi_\text{eff}(0)\text{ }(\text{V})$\footnote{From Table~\ref{HWPots}}&$\frac{d\phi_\text{eff}}{dq}\text{ }(\frac{\text{V}}{e})$\footnote{From least squares fit to $\phi_\text{eff}$} &$R_\text{Born}\text{}\left(\text{\AA}\right)$\footnote{From Eq.~\ref{BornRad}}\\ \hline
\ \ \ \ $+$ &2.0 &0.19&-11.1&1.28 \\
\ \ \ \ $+$& 2.6 & 0.04&-7.7&1.84\\
\ \ \ \ $-$& 2.6 & -0.04&-11.8&1.20\\
\hline
\end{tabular*}
\label{SEfits}
\end{threeparttable}
\end{table}

One remarkable result of this work is that DFT-MD calculations exhibit a very large CHA. In other words a negative charged hard sphere the size of fluoride has a `real' solvation free energy that is more than $200$ kJmol$^{-1}$ more negative than a cation of the same size (similar to potassium). This difference is dramatically larger than the experimental estimate of the CHA, where the fluoride anion has a solvation free energy that is  $\approx$ 50-100 kJmol$^{-1}$  more negative than potassium's. It is also much larger than the estimates of this quantity based on classical-MD where it is $\approx$100 kJmol$^{-1}$\cite{Hummer1996,Rajamani2004} after correcting for the surface potentials. 

This large difference is not due to a deficiency in the DFT-MD calculations. Rather, it serves as a harbinger that real quantum mechanical ions are very different to the idealized charged hard spheres treated here. A reason that the CHA is over-estimated is that charged hard spheres contain only electrostatics and an infinite repulsion. Other forms of interaction play an important role as well. For instance the exchange repulsion will significantly compensate for the  electrostatic CHA. Symmetry Adapted Perturbation Theory (SAPT)  calculations by \citeauthor{Pollard2016}\cite{Pollard2016}  make this clear, showing that the exchange interaction energy with the surrounding water molecules is approximately three times larger for fluoride than it is for potassium. Other terms such as dispersion \cite{Duignan2014a} and induction also need to be included for an accurate accounting of this effect. 


This large CHA has significant implications for classical-MD and continuum solvent models of ions in solution. Out findings suggest that in order to correctly reproduce the contributions to the solvation free energy these models should include a very large CHA  in addition to a large compensating charge dependent exchange energy. This picture is dramatically different to current classical-MD results where almost all of the CHA is contained in the electrostatic term and the exchange and dispersion energies are meant to be captured with a Lennard Jones interaction, which is relatively charge symmetric for a given ion size. Clearly classical-MD is not accurately reproducing the underlying physics. Having said that empirical models are phenomenological, \emph{i.e.}, they are designed to capture the experimental results in a simple way, rather than reproduce all the underlying physics. 

A similar critique of the continuum solvent model of Ref.~\citenum{Duignan2013a} can be made where it was \emph{assumed} that the electrostatic energy is charge symmetric and that all of the CHA can be explained with the dispersion interaction. This is clearly not the case as the simulations presented here show; charged hard spheres do show a dramatic solvation asymmetry. Interestingly though, the model seems to work fairly well for cations. The charging process is linear and this linearity allows us to extract values for the Born Radii from Eq.~\ref{BornRad} as shown in Table~\ref{SEfits}.  If the definition for the Born radii used in Ref.~\citenum{Duignan2013a} is used ($R_\text{Born}=R_\text{S}-0.84 \text{ \AA}$)  then an alternative derivation of the Born radii is possible. Values of  1.28\ \AA\ and 1.88\ \AA\ are predicted for the 2\ \AA\ and\ 2.6\ \AA\ ions respectively. (The peak in the ion-oxygen RDF is slightly larger than the hard sphere repulsion size). For the cations these values compare very well with the values given in Table~\ref{SEfits}. For the anion however the cavity size differs dramatically from this simple prediction.  There are many alternative definitions for the Born radius that have been proposed in the literature, many of which give different definitions for cations and anions. However, to the best of our knowledge none are conistent with the very small Born radius for the fluoride sized anion determined here, which remarkably, is even smaller than the crystal size of fluoride.  This indicates that a Born model may be a reasonable approximation for positive hard spheres but not for negative ones. This is consistent with the non-linear behavior of the anion at small charge states. The SAPT results presented by  \citeauthor{Pollard2016}\cite{Pollard2016} indicate that the asymmetry from the exchange term approximately cancels the asymmetry from the electrostatic and induction terms. This does provide some  justification for the continuum solvent model developed in Ref.~\citenum{Duignan2013a} where all of the asymmetry is assumed to arise from the dispersion interaction. 


It is important to note that the research presented in this study only examines
charged hard spheres and so it is not possible to directly determine the
solvation free energies of real ions. We have addressed this in Ref.~\citenum{Duignan2017a}. However, we have determined values
for the different definitions of the surface potentials. It is these surface
potentials that determine the conversion
between the different definitions of solvation free energy.  A central quantity
for the conversion between different definitions of the solvation free
energy is the dipolar potential due to a distant air-water interface, $\phi_\text{D}$.
\citeauthor{Hunenberger2011}\cite{Hunenberger2011} argue that the dipolar surface potential of the air-water 
interface ($\phi_\text{D}$) is approximately $+0.13$ V. There is a very large uncertainty in this estimate as it 
is determined primarily from indirect experimental methods that do not necessarily distinguish between the dipolar
potential, the Bethe potential, and the net potential. In addition, as discussed above this quantity depends 
on the choice of the origin of the water molecule and so it is not at all clear what choice is implicit 
in the experimental methods.
The only way to determine $\phi_\text{D}$ directly is with DFT-MD. Ref.~\citenum{Remsing2014} estimates this quantity using the oxygen atom as the center with BLYP-D2 and revPBE-D3. Both functionals arrive at a value of +0.48 V for $\phi_\text{D}$. Independent researchers\cite{Sulpizi2013} have also derived an almost identical value (0.47 V) using a significantly smaller box
(128 water molecules versus 340) and different basis sets. 
This consistency suggests that $\phi_\text{D}$ is being correctly estimated by DFT-MD simulations. The generalized gradient approximation (GGA) functionals used in the present study are approximate and so the values need to be confirmed by comparison with experimental measurements and higher level theoretical methods.\cite{DelBen2015}
The value for the TIP4P-Ew and TIP4P water models are quite similar ($\approx 0.4$ V to 0.6 V) \cite{Cendagorta2015,Warren2007a} indicating that this result is consistent with at least some classical-MD models. The MB-pol water model\cite{Medders2014} gives a value of approximately 0.3 V for the dipolar surface potential. This model reproduces the sum-frequency generation (SFG) spectra of the air-water interface\cite{Medders2016} which is sensitive to the water orientation and so this indicates that the GGA functionals may be overestimating the dipolar surface potential somewhat. 


   Last we determine the conversion of our `real' single ion free energies as obtained
   with Eq.~\ref{SEpart} to the bulk solvation free energies of Born theory. This conversion is given by the net potential. 
   It is important to note that the bulk values are inherently ion size and
   repulsion type specific.  Ref.~\citenum{Remsing2014}, corroborated by the results outlined here, 
   shows that the net potential is not overly sensitive to the size or nature of the cavity and that it tends to 
   lie between 0 and 0.2 V.  One possibility is that the 0.13 V value given in Ref.~\citenum{Hunenberger2011} is actually the net potential not the dipolar potential. The net potential does not appear to dramatically depend on the 
   cavity size and so the concept of a bulk solvation free energy still remains valid.
   Nevertheless, there is no unambiguous way to define an exact value for this type of solvation 
   free energy for all ions.  Because the Born free energies are best compared with the Bulk solvation 
   free energies, this implies that a size and repulsion specific correction is required to correct the 
   Born model to allow for comparison with `real' solvation free energies.  \citeauthor{Shi2013} \cite{Shi2013}  argue that a cavity size  of 6.15 \AA \ should be used to determine the net potential.  This is the size at which the cavity formation and Born solvation free energies cancel for monovalent ions. It should be noted that this is dramatically larger than any of the alkali-halide ions. 
   
  Beck and co workers also argue that a value of  $-0.4$ V for the net potential should be adopted.\cite{Beck2013,Shi2013,Pollard2014,Pollard2014a,Pollard2016}  This is based on multiple indirect lines of evidence based on both theory and experiment.  Firstly,  Ref.~\citenum{Marcus1987,Asthagiri2003} and \citenum{Ashbaugh2008} are used to support a value of $-$1065 kJmol$^{-1}$ for the bulk solvation free energies. Secondly, Ref.~\citenum{Pollard2014a} and Ref.~\citenum{Pollard2014}  are used to justify  a real  solvation free energies of close to $-$1105 kJmol$^{-1}$, which is similar to the CPA value. The difference between these values is used to infer a net potential of $\approx -0.4$ V. 
   
    Our work here, along with Ref.~\citenum{Remsing2014}, have provided estimates of this net potential 
    based in the framework of quantum mechanical simulation, and these estimates do not support a value of $-0.4$ V. 
    Our research instead supports the notion that 
    the net potential should be considered to make only a small contribution to the solvation free 
    energies, somewhere on the order of $\approx 0.1 \text{ V} \approx 10 \text{ kJmol}^{-1}$, namely bulk solvation free 
    energies  should be regarded as being close to the `real' values. 
    Although experimental evidence can in principle provide an indication of the surface potentials, interpretation of experiment to infer these surface potentials is generally very challenging and subjective. 
    
The values for the single ion solvation free energies determined by \citeauthor{tissandier1998}\cite{tissandier1998} on the basis of the CPA are considered by many to be the benchmark.\cite{camaioni2005}  This has recently been disputed however.\cite{Vlcek2013,Pollard2014,Vlcek2015} 

\section{Conclusions and Outlook}
In conclusion, we have outlined the simulation protocol
necessary to compute the electrostatic solvation free energy of 
charged hard spheres in water using DFT-MD. We have defined four types of single ion solvation 
free energy commonly used in the literature and outlined a prescription to convert
between them, linking to other author's notations and definitions where necessary. We have also provided best estimates for the values necessary to make these conversions. In particular, the net potential required to convert `real' to bulk solvation free energies is shown to be a small. ($\approx 0.1$ V) The dipolar surface potential necessary to convert `real' to intrinsic solvation free energies is shown to be 0.48 V.
Moreover, we can  correct the solvation
free energies calculated with standard implementations of Ewald summation in order to account for the unphysical electrostatic reference and arrive at physically reasonable values for the single ion solvation free energy. The path is now clear to calculate the solvation free energies of real ions, which is presented in Ref.~\citenum{Duignan2017a}. 

Our research also investigated the connection of Born theory to DFT-MD and found that the charging free energies of monovalent cations is consistent with linear response.  In contrast, non-linear charging behavior appears to exist for small anions at low charges. The Born model and classical-MD do not properly reproduce the  electrostatic solvation free energy of charged hard spheres and so should be considered  phenomenological approaches to ion hydration. A highlight of our research suggests that with DFT-MD the CHA is significantly larger for charged hard spheres than both the experimental estimates for real ions and for models of ions that use classical empirical interaction potentials. This result highlights the importance of local exchange and dispersion contributions to CHA that need to be incorporated into reduced models in order to move beyond phenomenology and capture the correct balance of the essential physics for ion solvation. 

\section{Supplementary Material}
See supplementary material for additional technical information regarding the simulation protocol.

\section{Acknowledgements}
We would like to thank Thomas Beck, Shawn Kathmann, Philippe H\"{u}enberger, Richard Remsing and John Weeks for helpful discussions. 
Computing resources were generously allocated by PNNL’s Institutional Computing program. 
This research also used resources of the National Energy Research Scientific Computing Center, a DOE Office of Science User Facility supported by the Office of Science of the U.S. Department of Energy under Contract No. DE-AC02-05CH11231. 
TTD, GKS and CJM were supported by the U.S. Department of Energy, Office of Science, Office of Basic Energy Sciences, Division of Chemical Sciences, Geosciences, and Biosciences. MDB was supported by MS$^{3}$ (Materials Synthesis and Simulation Across Scales) Initiative, a Laboratory Directed Research and Development Program at Pacific Northwest National Laboratory (PNNL).  PNNL is a multiprogram national laboratory operated by Battelle for the U.S. Department of Energy.

\bibliography{libraryabrev.bib}
\section{Supplementary Information}
\subsection{Chemical potential}
\label{ChemPotSec}
 The chemical potential of an ion $X$ in solution is given\cite{Ben-Naim1978} by:
\begin{equation}
\begin{split}
\mu^{\text{S}}_{X}&=k_{\text{B}}T\ln\rho_{X}^S\Lambda_{X}^3-k_{\text{B}}T\ln\left<e^{-\beta U_{XS}}\right>_{0}\
\end{split}
\end{equation}
where 
\begin{equation}
\begin{split}
-k_{\text{B}}T\ln\left<e^{-\beta U_{XS}}\right>_{0}=-k_{\text{B}}T\ln\frac{\int e^{-\beta U_{XS}} e ^{-\beta U_{N_s}}d\bm{R}^{N_s}}{\int e ^{-\beta U_{N_s}}d\bm{R}^{N_S}}
\end{split}
\end{equation}
where $d\bm{R}^{N_S}$ indicates that the integral is over all configurations of the solvent molecules and  $U_{N_s}$ and $U_{XS}$ are defined in text. 

The chemical potential of the ion in vapor infinitely far from the air-water interface is given by: 
\begin{equation}
\mu^{\text{V}}_{X}=k_{\text{B}}T\ln\rho_{X}^V\Lambda_{X}^3+E^\text{Vac}_X
\end{equation} 
The zero of the electrostatic potential must be the same for both the the vapor and condensed phase calculations. The most natural choice is that the electrostatic potential in vapor infinitely far away from the air-water interface should be zero. This means that the only contribution to  $E^\text{Vac}_X$ is the electronic energy of the ion in vacuum.
For a classical water model or a point charge this energy is zero but not for a real ion with an electronic wave function. This choice of the zero of the electrostatic potential does mean that the potential created by the bulk air-water interface will contribute to the solvation free energy of the ion as it will alter the potential inside the water where the ion resides. 

The solvation free energy is given by: 
\begin{equation}
\Delta\mu_{X}=\mu^{\text{S}}_{X}-\mu^{\text{V}}_{X}=k_{\text{B}}T\ln\frac{\rho_{X}^{S}}{\rho_{X}^{V}}-k_{\text{B}}T\ln\left< e ^{-\beta U_{XS}}\right>_{0}-E^\text{Vac}_X
\end{equation}
where the last two terms comprise the excess chemical potential\cite{Ben-Amotz2005a,Beck2006} or the point to point solvation free energies\cite{Hunenberger2011} or the local standard transfer free energy. \cite{Ben-Naim1978} They correspond to the solvation free energy when the standard state concentration of ions in solution is the same as in vapor. Often this is stated to be one Molar. Many researchers use a standard state of an ideal gas at 1 atm for the concentration of ions in the vapor and 1 M for the ions in solution, which is confusing and unnecessary.\cite{Ben-Naim1978} We therefore use the definition above throughout and adjust other researchers's values accordingly. With this standard state we arrive at Eq.~1 in the main text.

\subsection{Point charge and cavity partitioning}
\label{PointChSec}
The following expression demonstrates how to break the solvation free energy into a cavity formation free energy and a  charging free energy:
\begin{equation}
\begin{split}
\mu^*_{X}&=-k_{\text{B}}T\ln \left< e ^{-\beta U_{XS}}\right>_{0}=-k_{\text{B}}T\ln\frac{\int e ^{-\beta(U_{XS}+U_{N_s})}d\bm{R}^{N_S}}{\int e ^{-\beta U_{N_s}}d\bm{R}^{N_S}}\\&=-k_{\text{B}}T\ln\frac{\int e ^{-\beta(U_{\text{PC}}+ U_{\text{\text{Cav}}}+U_{N_s})}d\bm{R}^{N_S}}{\int e ^{-\beta U_{N_s}}d\bm{R}^{N_S}} -k_{\text{B}}T\ln\frac{\int e ^{-\beta(U_{\text{\text{Cav}}}+U_{N_s})}d\bm{R}^{N_S}}{\int e ^{-\beta(U_{\text{\text{Cav}}}+U_{N_s})}d\bm{R}^{N_S}}\\&=
-k_{\text{B}}T\ln\frac{\int e ^{-\beta(U_{\text{\text{Cav}}}+U_{N_s})}d\bm{R}^{N_S}}{\int e ^{-\beta U_{N_s}}d\bm{R}^{N_S}}-k_{\text{B}}T\ln\frac{\int e ^{-\beta(U_{\text{PC}}+ U_{\text{\text{Cav}}}+U_{N_s})}d\bm{R}^{N_S}}{\int e ^{-\beta(U_{\text{\text{Cav}}}+U_{N_s})}d\bm{R}^{N_S}}\\ &=-k_{\text{B}}T\ln \left< e ^{-\beta U_{\text{\text{Cav}}}}\right>_{0}-k_{\text{B}}T\ln\left< e ^{-\beta U_{\text{PC}}}\right>_{U_{\text{\text{Cav}}}}=\mu^*_{\text{Cav}}+\mu^*_{\text{PC}}
\end{split}
\label{cavSEdef}
\end{equation}
 \subsection{Charging energy}
\label{ChESec}
We can slowly charge the ion up in small increments in order to calculate the solvation free energy accurately and to test how linear this is. The resulting expression is: 
\begin{equation}
\mu^*_{\text{PC}}=\sum_{i=0}^{N-1}-k_{\text{B}} T\ln\left< e ^{-\beta \big(U_{\text{PC}}((i+1)\Delta q)- U_{\text{PC}}\left(i\Delta q\right)\big)}\right>_{U_{\text{Cav}}+U_\text{PC}(i\Delta q)}
\label{PCuSE}
\end{equation}
where $\Delta q=q/N$.
Similarly, we can also turn the charge off in small increments and determine a relatively independent estimate of this quantity using the inverse expression:
\begin{equation}
\mu^*_{\text{PC}}=\sum_{i=1}^{N}k_{\text{B}} T\ln\left< e ^{\beta \big(U_{\text{PC}}(i\Delta q)- U_{\text{PC}}\left(\left(i-1\right)\Delta q\right)\big)}\right>_{U_{\text{Cav}}+U_\text{PC}(i\Delta q)}
\label{PCdSE}
\end{equation}
If we assume that the fluctuations are Gaussian  the integrals can be performed analytically,\cite{Beck2006}  arriving at the following expressions:
\begin{equation}
\begin{split}
\mu^*_{\text{PC}}=&\sum_{i=0}^{N-1}\Big< (U_{\text{PC}}((i+1)\Delta q)- U_{\text{PC}}(i\Delta q))\Big>_{U_{\text{Cav}}+U_\text{PC}(i\Delta q)}\\&-\frac{1}{2k_{\text{B}}T}\Big< \delta\left[U_{PC}((i+1)\Delta q)-U_{PC}(i\Delta q)\right]^2\Big>_{U_{\text{\text{Cav}}}+U_{\text{\text{PC}}}(i\Delta q)}
\label{PCuSEG}
\end{split}
\end{equation}
and
\begin{equation}
\begin{split}
\mu^*_{\text{PC}}=&\sum_{i=1}^{N}\Big< (U_{\text{PC}}(i\Delta q)- U_{\text{PC}}((i-1)\Delta q))\Big>_{U_{\text{Cav}}+U_\text{PC}(i\Delta q)}\\&+\frac{1}{2k_{\text{B}}T}\Big< \delta\left[U_{PC}(i\Delta q)-U_{PC}((i-1)\Delta q)\right]^2\Big>_{U_{\text{\text{Cav}}}+U_{\text{\text{PC}}}(i\Delta q)}
\label{PCdSEG}
\end{split}
\end{equation}
where the $\delta[U]^2$ notation indicates the square of the standard deviation. The assumption of Gaussian fluctuations does not significantly alter the results. This does not necessarily prove that the fluctuations are Gaussian however as such small simulation times can only be expected to reliably estimate the first and second cumulants of the probability distribution. Rather the agreement between the charging and decharging steps should indicate whether the Gaussian approximation is acceptable. For the anion we use a smaller charge increment than for the cations in order to better probe the non-linearity.

\subsection{Origin dependence of $\phi_{D}$ and $\phi_\text{B}$}
\label{OrigDepSec}
We can see that the partitioning of the surface potential of the air-water interface into a dipolar and quadrupolar trace contribution depends on the choice of origin with the following argument. First we write the dipolar surface potential explicitly in terms of an integral over the orientational distribution of the dipoles at the air-water interface.
\begin{equation}
\phi_\text{D}=-\epsilon_0^{-1}\int_{z_l}^{z_v} dz P_z(z)=-\frac{1}{4\pi\epsilon_0}\int_{z_l}^{z_v} dz \int_0^\pi d\theta \int_0^{2\pi}d\phi \sin \theta P(z,\theta)\mu\cos\theta
\end{equation}
where $ P(z,\theta)$ is the density of dipoles with a given position and orientation and $\mu$ is the molecular dipole moment. $\theta$ gives the orientation of the dipole moment relative to the $z$ axis, which is normal to the interface.  $z_l$ and $z_v$ are points deep in the liquid and vapor phase respectively. 
If we now change the origin of the water molecule, $\mu$ will not change as the dipolar moment of a neutral molecule is independent of the choice of the origin. However, the distribution of the dipole moment will change. In particular, if we move the center of the water molecule by a distance $\boldsymbol{d}$, the dipolar surface potential becomes:

\begin{equation}
\begin{split}
\phi_\text{D}(\boldsymbol{d})=&-\frac{1}{4\pi\epsilon_0}\int_{z_l}^{z_v} dz \int_0^\pi d\theta \int_0^{2\pi}d\phi P(z-\left(\boldsymbol{\hat{\mu}}\cdot\boldsymbol{d}\right)\cos\theta,\theta)\mu\cos\theta \sin \theta
\\=&-\frac{\mu}{2\epsilon_0} \int_0^\pi d\theta  \int_{z_l}^{z_v} dzP(z-\left(\boldsymbol{\hat{\mu}}\cdot\boldsymbol{d}\right)\cos\theta,\theta)\cos\theta \sin \theta\\=&-\frac{\mu}{2\epsilon_0} \int_0^\pi d\theta  \cos\theta \sin \theta\int_{z_l-\left(\boldsymbol{\hat{\mu}}\cdot\boldsymbol{d}\right)\cos\theta}^{z_v-\left(\boldsymbol{\hat{\mu}}\cdot\boldsymbol{d}\right)\cos\theta} duP(u,\theta)\\=&-\frac{\mu}{2\epsilon_0} \int_0^\pi d\theta  \cos\theta \sin \theta\left(\int_{z_l}^{z_v} duP(u,\theta)+\rho_w\left(\boldsymbol{\hat{\mu}}\cdot\boldsymbol{d}\right)\cos\theta\right)\\=&\phi_\text{D}(\boldsymbol{0})-\frac{\rho_w\left(\boldsymbol{\mu}\cdot\boldsymbol{d}\right)}{2\epsilon_0} \int_0^\pi d\theta  \cos^2\theta \sin \theta=\phi_\text{D}(\boldsymbol{0})-\frac{\rho_w\left(\boldsymbol{\mu}\cdot\boldsymbol{d}\right)}{3\epsilon_0} 
\end{split}
\end{equation}
where we have taken advantage of the fact that the dipole density goes to 0 around $z_v$ and $\rho_w$ around $z_l$.
From Eq.~6 in the main text it is trivial to show that the Bethe potential change with change in the position of the origin is given by:
\begin{equation}
\phi_\text{B}(\boldsymbol{d})=\phi_\text{B}(\boldsymbol{0})+\frac{\rho_w\left(\boldsymbol{\mu}\cdot\boldsymbol{d}\right)}{3\epsilon_0} 
 \end{equation}
$\phi_\text{B}(\boldsymbol{d}) +\phi_\text{C}(\boldsymbol{d})$ is obviously independent of $\boldsymbol{d}$.

For a simple example let us assume that every molecule is simply a positive and negative charge separated by some distance $l$  with an isotropic Heaviside step function distribution, \emph{i.e.},  we assume the center of the molecule is in between the two charges  and that the molecules are isotropically distributed about this center and that the interface is infinitely sharp and located at $z_0$. The dipolar surface potential for this interface is obviously 0.
\begin{equation}
\phi_\text{D}=-\epsilon_0^{-1}\int_{z_l}^{z_v} dz P_z(z)=-\frac{1}{4\pi\epsilon_0}\int_{z_l}^{z_v} dz \int_0^\pi d\theta \int_0^{2\pi}d\phi \rho_w H(z_0-z)lq\cos\theta\sin \theta=0
\end{equation}
If we adjust the origin of the water molecule by a distance $l/2$ so that the origin is now on top of the negative charge then the new dipolar surface potential can be written as:
\begin{equation}
\phi_\text{D}=-\frac{1}{4\pi\epsilon_0}\int_{z_l}^{z_v} dz \int_0^\pi d\theta \int_0^{2\pi}d\phi   \rho_w H(z_0-z-l/2\cos\theta)lq\cos\theta \sin \theta
\end{equation}
This integral can be written as: 
\begin{equation}
\phi_\text{D}=\frac{-lq \rho_w}{2\epsilon_0}\int_{z_0-\frac{l}{2}}^{z_0+\frac{l}{2}} dz \int_{\cos^{-1}\left(\frac{2(z-z_0)}{l}\right)}^\pi d\theta \cos\theta \sin \theta
\end{equation}
which can be performed analytically to arrive at the following expression for the dipolar potential with a shifted origin:
\begin{equation}
\phi_\text{D}=\frac{l^2q \rho_w }{6\epsilon_0}
\end{equation}
From Eq.~6 in the main text we can see that the Bethe potential will shift from 0 to the following expression:
\begin{equation}
\phi_\text{B}=-\frac{l^2q \rho_w }{6\epsilon_0}
\end{equation} 
which precisely cancels the change in the dipolar surface potential, showing that the partitioning is dependent of the choice of the origin but that the total surface potential is not. This  must be the case as the total surface potential can be calculated without assuming a molecular center at all. 

 The origin dependence of the two contributions to the surface potential is somewhat counter-intuitive as it is not immediately clear why the dipolar contribution should change with a different choice for the origin given that the dipole moment of a neutral molecule is independent of the choice of the origin. The reason for this dependence on the choice of the origin can be made intuitively clear however. The magnitude of the dipole moment of each molecule doesn't depend on the choice of the origin, but the spatial and orientational distribution of those dipoles does depend on where the origin is chosen to be. If the negative  charge is chosen to be the center of the molecule then the molecules will no longer be isotropically oriented on average. Instead they will be more likely to point into water on average.  This has important implications for spectroscopic studies of the air-water interface, which draw conclusions about the average orientation of water molecules at the air-water interface because it highlights that care needs to be taken to be certain to determine what molecular center is being assumed.\cite{Byrnes2011} 

\subsection{Surface potential values}
\label{SurfPotSec}
There are  two contributions to the Bethe potential. The first is from the combination of the hydrogen charges and the electron density, this corresponds to the value given in Ref.~\citenum{Remsing2014} and is a real physical quantity that can be measured experimentally.\cite{Kathmann2011} However, the value given in Ref.~\citenum{Remsing2014} will not totally agree with experiment as two of the oxygen atom's inner-electrons are treated with a pseudopotential. This difference can be effectively corrected with an all electron calculation on a single water molecule. The second contribution, as described in the SI of Ref.~\citenum{Remsing2014} is from the spread of the pseudopotentials used to model the positive charges at the center of the atoms. This contribution is not  physically real as the real atomic cores have a much more sharply defined density and is not included in the values given in  Ref.~\citenum{Remsing2014}, but it does need to be included in correcting the Ewald solvation free energies and so we include it here. 

The Bethe potential of water with a solute present is distinct from the Bethe potential of pure water. The given definition of the Bethe potential accounts for this (Eq.~6 in the main text) as it uses the box size rather than relying on the density of bulk water. This is more accurate than the alternative definition given by \citeauthor{Hunenberger2011}\cite{Hunenberger2011}, which uses the Bethe potential of bulk water  and then includes an approximate correction to account for the presence of the cavity that relies on an ambiguous cavity size. Because the Bethe potential is significantly larger for DFT-MD calculations compared to classical-MD, it is particularly important to estimate this correction accurately. This correction for the Bethe potential is referred to as the C1 correction in Ref.~\citenum{Kastenholz2006}.  

The Bethe potential was calculated for each situation to examine how much it varies with cavity size and charge and the results are presented in Table~\ref{BethePots}.
\begin{table}
\begin{threeparttable}
\centering
\caption[]{Bethe potentials ($\phi_\text{B}$)}
 \begin{tabular*}{1\textwidth}{@{\extracolsep{\fill}}clclclclc}\hline
Functional &Cavity size(\AA)&Charge(e)&$\phi_\text{B}(V)$ \\ \hline
revPBE-D3& 2.0 & $0$&3.415\\
revPBE-D3& 2.0 & $1$&3.409\\
revPBE-D3& 2.6 & $0$&3.414\\
revPBE-D3& 2.6 & $-1$&3.402\\
BLYP-D2& 2.0& $0$&3.417 \\
\hline
\end{tabular*}
\label{BethePots}
\end{threeparttable}
\end{table}

\subsection{Ewald correction term}
\label{EwCorrSec}
The $\mu_\text{Ew-Corr}$ correction is made up of two largely compensating contributions. One contribution is from the self energy of the ion due to the interaction with its compensating background (gellium) and image charges.  The second contribution is from the finite size of the box, which limits the number of water molecules the ions can polarize. Due to the long range of the Coulomb interaction this correction is significant even for moderately large boxes. Fortunately this long-range correction for finite size effect depends mainly on the dielectric constant of the solution not on any molecular details and  \citeauthor{Hunenberger1999}\cite{Hunenberger1999} have derived it rigorously using continuum theory.\cite{Hunenberger1999} 

These two corrections combine to give:
\begin{equation}
\mu_\text{Ew-Corr}= \frac{q_I^2}{8\pi\epsilon_0L}\xi_{\text{Ew}}-\frac{q_I^2}{8\pi\epsilon_0L}\left(1-\frac{1}{\epsilon_w}\right)\left(\xi_{\text{Ew}}+\frac{4\pi}{3}\left(\frac{R_I}{L}\right)^2 -\frac{16\pi^2}{45}\left(\frac{R_I}{L}\right)^5 \right)
\label{Ew-corr}
\end{equation}
where $\xi_\text{Ew}=-2.837297$ is the Wigner constant, which is associated with the energy of charging an infinite cubic array of point charges surrounded by a diffuse compensating charge calculated with Ewald summation.   This expression is the sum of 
Eq.~(35) and Eq.~(39) from Ref.~\citenum{Hunenberger1999}. The second term in this expression is referred to as the B type correction by \citeauthor{Kastenholz2006}.\cite{Kastenholz2006} $R_I$ corresponds to the size of the ion. $L$ is the length of  one side of the box.  These two corrections, which are quite large on their own, mostly cancel for water in reasonably sized boxes.  This explains why previous work has not needed to include these corrections.\cite{Grossfield2003}   There is an additional boundary condition associated with Ewald summation, which is that the electric field averages to zero over the box.\cite{Kastenholz2006} This is what is referred to as tinfoil boundary conditions. This appears to be a reasonable approximation for ion solvation in water due to its high dielectric constant and therefore we do not correct for it.   There is also a correction associated with the effect the electrostatic potential inside the cavity has on the average potential. This is referred to as the C2 correction, but it is very small and is neglected here. \cite{Kastenholz2006} 
These expressions are derived and tested by \citeauthor{Kastenholz2006}\cite{Kastenholz2006}  by comparing  different methodologies and using very large water clusters, up to 17,454 waters, to show that they are accurate.

\subsection{Effective potential}
\label{EffPotSec}
Normally the point charge solvation energy is rewritten using Thermodynamic Integration (TI) as an integral of the potential as a function of charge. However, we cannot use this expression here as there is a polarization response of the water, and so the point charge interaction energy is not simple $q\phi$, but actually:
\begin{equation}
U_{\text{PC}}=q\phi_0+\frac{q\phi_I(q)}{2}
\end{equation}
where $\phi_0$ is the electrostatic potential created when there are no ions there and $\phi_I(q)$ is the change in the potential due to the electronic polarization of the water molecules due to the presence of the ion. Ref.~\citenum{Israelachvili2011} shows that there is a factor of half because half of the induction energy is used to polarize the water molecules. The point charge solvation free energy is therefore:
\begin{equation}
\mu^*_{\text{PC}}=-k_{\text{B}} T\ln\left< e ^{-\beta U_{\text{PC}}}\right>_{U_{\text{Cav}}}=-k_{\text{B}}T\ln\frac{\int e ^{-\beta(q\phi_0+\frac{q\phi_I(q)}{2}+ U_{\text{\text{Cav}}}+U_{N_s})}d\bm{R}^{N_s}}{\int e ^{-\beta(U_{\text{\text{Cav}}}+U_{N_s})}d\bm{R}^{N_s}}
\end{equation}
and the derivative with respect to q is:
\begin{equation}
\frac{d\mu^*_{\text{PC}}}{dq}= \frac{ \int \left(\phi_0+\frac{\phi_I(q)}{2}+\frac{q}{2}\frac{d\phi_I(q)}{d q}  \right)  e ^{-\beta(q\phi_0+\frac{q\phi_I(q)}{2}+ U_{\text{\text{Cav}}}+U_{N_s})}d\bm{R}^{N_s}}{\int e ^{-\beta(q\phi_0+\frac{q\phi_I(q)}{2}+ U_{\text{\text{Cav}}}+U_{N_s})}d\bm{R}^{N_s}}
\label{dmubdq}
\end{equation} 
We can therefore write the solvation free energy as:
\begin{equation}
\mu^*_{\text{PC}}=\int_{0}^Q\frac{d\mu^*_{\text{PC}}}{dq}=\int_{0}^{Q}dq\left<\phi_0+\frac{\phi_I(q)}{2}+\frac{q}{2}\frac{d\phi_I(q)}{d q} \right>_{U_{\text{Cav}}+U_\text{q}}
\label{phiint}
\end{equation}
This expression converges to the normal integration of the electrostatic potential in the case of non-polarizable water. The expression is clearly significantly more complex for an electronically polarizable solvent and so simply using the total energies is more straightforward than using the potential. 
We can estimate the effective potential with the following expression:
\begin{equation}
 \phi_\text{eff}(q)=\frac{d\mu^*_{X}}{dq}=\frac{d\mu^*_\text{PC}}{dq}\approx\frac{\mu^*_{X}(q+\Delta q /2)-\mu^*_{X}(q-\Delta q /2)}{\Delta q}
\end{equation}

\subsection{Distributions}
Here we show some examples of the energy and potential distributions with fitted Gaussians. These plots show that the distributions are generally Gaussian.  The large CHA means that  there must be some non Gaussian behavior in the tails but the distributions in these regions are not well converged.
\begin{figure}
                \includegraphics[width=1\textwidth]{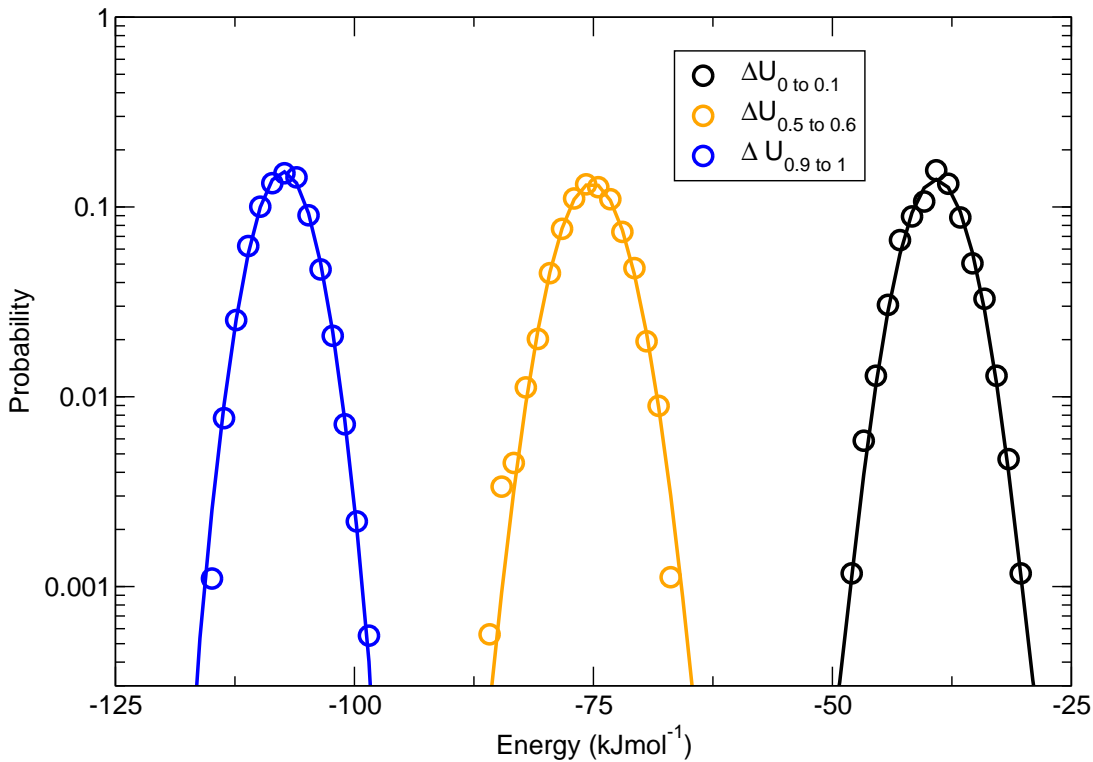}
             \caption[]{Distributions of the  energy change for the 2.6 \AA\ sized cation as the charge it turned on up to 1 $e$ in units of 0.1 $e$ increments. This is the first term in Eq.~\ref{PCuSEG}. Gaussian fits are shown with lines. }
        \label{Chargingdistributions}
\end{figure}

\begin{figure}
                \includegraphics[width=1\textwidth]{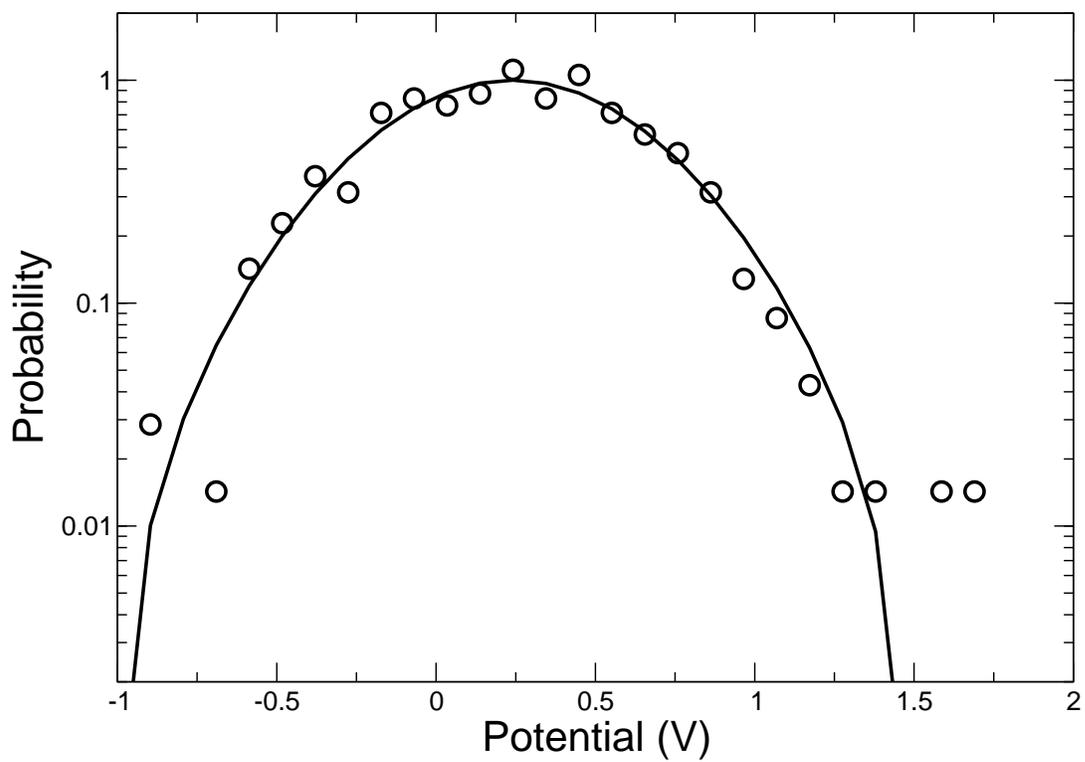}
             \caption[]{Distributions of net potential at the center of a 2 \AA\ uncharged cavity.  }
        \label{Chargingdistributions}
\end{figure}

\end{document}